\documentclass[aps,pre,superscriptaddress,bibtex,twocolumn,longbibliography]{revtex4-1}

\usepackage[utf8]{inputenc}
\usepackage[T1]{fontenc}

\usepackage[tbtags]{amsmath}
\usepackage{amssymb}
\usepackage{mathtools}
\usepackage{bm}

\usepackage{float}

\usepackage[section]{placeins}
\usepackage[margin=1in]{geometry}

\usepackage[textsize=scriptsize,]{todonotes}
\makeatletter \def\@captype{figure} \makeatother
\usepackage{xcolor}

\usepackage[colorlinks=true,urlcolor=blue,linkcolor=blue,citecolor=blue]{hyperref}

\newcommand{\be}{\begin{equation}}
\newcommand{\ee}{\end{equation}}
\newcommand{\ben}{\begin{eqnarray}}
\newcommand{\een}{\end{eqnarray}}
\newcommand{\la}{\langle}
\newcommand{\ra}{\rangle}

\newcommand{\genmatDrate}{\genmatDrate_{\Doobletter}}

\begin{document}

\title{Taming nonlinear energy diffusion: The case of time-crystal energy condensates}
\author{P.I. Hurtado}
\email{phurtado@onsager.ugr.es}
\affiliation{Departamento de Electromagnetismo y F\'isica de la Materia, Universidad de Granada, Granada 18071, Spain}
\affiliation{Institute Carlos I for Theoretical and Computational Physics, Universidad de Granada, Granada 18071, Spain}
\author{G. Cort\'es-Guill\'en}
\affiliation{Departamento de Electromagnetismo y F\'isica de la Materia, Universidad de Granada, Granada 18071, Spain}
\affiliation{F\'isica Te\'orica, Universidad de Sevilla, Apartado de Correos 1065, E-41080 Sevilla, Spain}
\date{\today}

\begin{abstract}
We study a bulk-driven nonlinear variant of the Kipnis-Marchioro-Presutti model of stochastic energy diffusion in which local collisions are biased to induce a net energy flow, resembling the effect of an external field. Starting from the microscopic master equation, we derive the hydrodynamic description of the driven system via a local equilibrium approximation, obtaining explicit expressions for the energy current and the associated diffusivity and mobility transport coefficients, which are nonlinear functions of the local energy density. We test our findings in kinetic Monte Carlo simulations of the model and, as a proof of concept, we demonstrate the versatility of this driving mechanism to control nonlinear energy transport by inducing time-crystalline phases. In particular, we show that appropriately designed packing fields induce the spontaneous formation of traveling energy condensates, exhibiting robust long-range temporal order reminiscent of continuous time crystals. Our results provide a simple yet powerful framework to study bulk-driven nonlinear energy diffusion in stochastic many-body systems, offering a bridge between microscopic dynamics, macroscopic transport, and controlled spatiotemporal order.
\end{abstract}

\maketitle

\section{Introduction}
\label{sec1}

Simple models have long played a central role in theoretical physics \cite{anderson72a, peierls80a, goldenfeld99a, widom74a, wigner60a, lebowitz99b, baxter16a}. By stripping away microscopic complexity while retaining the essential ingredients of the phenomena under study, such models provide a controlled arena in which general mechanisms can be identified, analyzed, and often understood with a degree of clarity that would be impossible in more realistic settings. From the Ising model or the hard-sphere fluid in equilibrium statistical mechanics to exclusion processes in nonequilibrium physics, minimal models have repeatedly proven to be reliable tools for uncovering universal behavior, testing theoretical frameworks, and guiding the interpretation of experiments \cite{marro05a}. Their success lies precisely in their ability to isolate the core physical principles governing collective behavior, while remaining amenable to analytical treatment and efficient numerical investigation.

In the context of nonequilibrium statistical physics, one of the paradigmatic examples of such a minimal yet powerful framework is the Kipnis-Marchioro-Presutti (KMP) model of heat conduction \cite{kipnis82a}, a true bridge between microscopic dynamics and macroscopic transport. Introduced four decades ago, the KMP model consists of a one-dimensional ($1d$) lattice in which nearest-neighbor sites randomly exchange energy through stochastic collisions that conserve the total energy locally. Despite its simplicity, the model marked a milestone in the field by providing the first rigorous microscopic derivation of Fourier’s law of heat conduction \cite{bonetto00a, lepri03a, dhar08a, lepri16a}. Since then, the KMP model has proven instrumental for many modern developments in nonequilibrium physics. For instance, together with the symmetric simple exclusion process, it is one of the very few interacting many-body systems for which the nonequilibrium quasipotential is exactly known \cite{derrida98b, derrida01a, bertini01b, bertini05b}. The model has played a central role in the development of macroscopic fluctuation theory \cite{bertini15a}, serving as a benchmark system for the study of current fluctuations and large deviations \cite{touchette09a, bertini05a, bertini06a, derrida07a, hurtado14a, perez-espigares19a, hurtado25a}, including time-dependent \cite{derrida09a, krapivsky12a, meerson13a, bettelheim22a} and high-dimensional settings \cite{perez-espigares16a, tizon-escamilla17a, garrido19a}. In this context, the KMP model has been used to test the additivity principle for current fluctuations \cite{bodineau04a, hurtado09c, hurtado10a}, i.e. the conjecture that long-time current fluctuations in $1d$ diffusive systems are often dominated by mostly time-independent optimal trajectories, providing also one of the first clear demonstrations of spontaneous symmetry breaking at the fluctuating level \cite{bodineau06a, hurtado11a, perez-espigares13a, zarfaty16a}. The KMP model has also served as a natural testing ground for fluctuation theorems \cite{evans93a, evans94a, gallavotti95a, gallavotti95b, kurchan98a, lebowitz99a, hurtado09c, hurtado10a, hurtado11b, perez-espigares15a} and duality properties in stochastic many-body systems \cite{giardina07a, giardina09a, carinci13a, tailleur07a}. Beyond its original formulation, the KMP idea has inspired a wide class of generalized transport models, including nonlinear versions to investigate compact wave propagation \cite{hurtado12a}, systems with multiple conserved quantities such as the kinetic exclusion process \cite{gutierrez-ariza19a}, or dissipative variants relevant to driven granular matter \cite{prados11a, prados12a, hurtado13a, lasanta15a, manacorda16a, plata16a}. 

Crucially, in most cases mentioned above the KMP model and its variants are driven out of equilibrium via boundary gradients. This boundary driving, though instrumental, severely constraints the range of accesible phenomena and the repertoire of control tools. In particular, boundary driving provides only indirect and essentially static control over the bulk dynamics. On the other hand, introducing a bulk driving mechanism by manipulating microscopic energy exchanges throughout the system overcomes these limitations. This opens a qualitatively new route to controlling nonlinear energy diffusion, enabling the design of targeted transport responses beyond stationary gradient-driven flows.

With this idea in mind, we study in this work a bulk-driven nonlinear generalization of the KMP model controlled by an external field. The microscopic KMP dynamics \cite{kipnis82a} is modified in two key ways. First, the collision kernel is generalized from the constant KMP rate to a generic energy-dependent kernel $\pi(\nu)$, with $\nu$ the total pair energy, thus allowing for nonlinear diffusive behavior already at the microscopic level. Second, the stochastic energy redistribution parameter in each local collision is not drawn from the uniform KMP distribution, but from an asymmetric field- and energy-dependent distribution, so that collisions remain strictly energy-conserving while becoming statistically biased and inducing a net energy current in the bulk. Starting from the exact master equation of the model, we obtain the associated balance equations, see Eqs. \eqref{continuityeq} and \eqref{balanceeq} below. We then obtain its hydrodynamic description by employing a local equilibrium approximation \cite{prados12a, gutierrez-ariza19a}. This approach allows us to obtain explicit expressions for the constitutive relation of the energy current, see Eq. \eqref{jcurrent}, as well as for the associated transport coefficients—namely the diffusivity \eqref{diff} and the mobility \eqref{mobi}—which acquire a nonlinear dependence on the local energy density through the microscopic collision rate. We then validate these results in kinetic Monte Carlo simulations of the microscopic model for representative choices of nonlinear collision kernels, including in particular power-law rates, measuring the current response to different constant external fields. To demonstrate the power and flexibility of this bulk-driving mechanism, we further use it to study the emergence of continuous time-crystal phases \cite{wilczek12a, zakrzewski12a, sacha18a, sacha20a} in energy diffusion. In particular, we implement an external packing field coupled to the energy field fluctuations, capable of triggering a second-order phase transition to a continuous time-crystal phase \cite{hurtado-gutierrez20a, hurtado-gutierrez23a, hurtado-gutierrez25a, hurtado-gutierrez25b} characterized by the formation of robust traveling energy condensates that break continuous time-translation symmetry. This shows how this bulk-driven mechanism can be used to control energy transport.

Although this work is primarily a proof-of-concept study on the interplay between local nonlinear diffusion and bulk driving in a fluctuating environment, extending the well-known KMP paradigm \cite{kipnis82a}, we expect our results to be relevant for a broad class of engineered diffusive media where local control of energy flow is possible. This is the case e.g. of thermal metamaterials or synthetic lattices based, for instance, on arrays of coupled microresonators or equivalent RC circuits, or engineered nanoplasmonic or photonic arrays, possibly combined with local feedback protocols and controlled noise sources. From this perspective, the main physical novelty of our work is not tied to a particular realization, but to showing that nonlinear energy diffusion can be actively shaped and controlled by a suitable bulk drive. This goes beyond the standard boundary-driven framework, and shows that bulk driving provides a qualitatively new route to engineer nonequilibrium energy flows and stabilize nontrivial transport states.

The paper is organized as follows. In Sec. II we define the driven nonlinear energy diffusion model and specify its microscopic stochastic dynamics. In Sec. III we derive the hydrodynamic equations governing the evolution of the macroscopic energy density field, obtaining explicit expressions for the current field and the transport coefficients. In Sec. IV we validate numerically our findings and explore the consequences of this driving mechanism. We first corroborate the hydrodynamic predictions through extensive numerical simulations under constant external fields. We then present a proof of concept by employing a packing field to induce the formation of time-crystalline energy condensates. Finally, Sec. V is devoted to a discussion of our results, their implications for controlled energy transport in nonequilibrium systems, and a summary of our conclusions.

\section{Bulk-driven nonlinear energy diffusion model}
\label{sec2}

The system is defined in a 1D lattice with $L$ sites. A configuration after the $n$-th event is given by $\bm{\epsilon}_n=\{\epsilon_{i,n}\}$, $i\in[1,L]$, where $\epsilon_{i,n} \geq 0$ is the \emph{energy} of the $i$-th site at step $n$. The dynamics is stochastic and evolves in continuous time through a sequence of instantaneous collision events between randomly selected nearest-neighbor pairs, whose occurrence times follow Poisson statistics. At each event, a pair $(i,i+1)$ --identified by the index $i$ of its leftmost lattice site-- is chosen with probability 
\[
\Pi(i|\bm{\epsilon}_n)\equiv \frac{\pi(\nu_{i,n})}{L \Omega(\bm{\epsilon}_n)} \, ,
\]
where $\pi(\cdot)$ is some non-negative function of the total pair energy $\nu_{i,n}\equiv \epsilon_{i,n}+\epsilon_{i+1,n}$, and the normalization factor $\Omega(\bm{\epsilon}_n)\equiv L^{-1}\sum_{l=1}^L \pi(\nu_{l,n})$ is defined so as to remain finite in the hydrodynamic limit $L\to \infty$. In this paper we work with periodic boundary conditions, so we identify sites $L+1\leftrightarrow 1$. In each collision, the total pair energy is redistributed stochastically between both sites, so
\begin{equation}
\epsilon_{i,n+1}=\alpha_n \nu_{i,n} \, , \quad \epsilon_{i+1,n+1}=(1-\alpha_n) \nu_{i,n} \,,
\label{collisionrule}
\end{equation}
with $\alpha_n\in [0,1]$ a random collision parameter drawn from a probability distribution function to be defined below. Once the collision is completed, the system clock is advanced by a random time increment, $\tau_{n+1}=\tau_n + \delta\tau_n/L$, with $\delta\tau_n$ drawn from an exponential distribution, 
\[
P(\delta\tau_n)=\frac{1}{\Omega(\bm{\epsilon}_n)} \textrm{e}^{-\delta\tau_n/\Omega(\bm{\epsilon}_n)} \, ,
\]
such that the typical time increment is the inverse of the total escape rate from the current configuration, $[L \Omega(\bm{\epsilon}_n)]^{-1}$.

In the original KMP version of this model, the random collision parameter $\alpha_n$ is homogeneously distributed in the unit interval, resulting in unbiased energy diffusion at the hydrodynamic level \cite{kipnis82a}. Similar choices were made in the nonlinear and dissipative generalizations of the KMP model introduced some time ago \cite{prados12a, hurtado12a, hurtado13a}. In order to drive energy diffusion in some preferred direction, mimicking the effect of a constant driving field $\chi$, we now tilt the probability distribution for $\alpha_n$ (see also \cite{carinci16a}). In particular, the random collision parameter is drawn from a distribution $f_\chi(\alpha_n|\nu_{i,n})$ that depends on the total pair energy $\nu_{i,n}$. The particular form of this distribution can be derived by imposing the local detailed balance condition on the transitions \cite{spohn12a}, so as to ensure that macroscopic fluctuations satisfy a fluctuation-dissipation relation locally \cite{bertini15a}. If we denote as $\bm{\epsilon}_{\alpha}^{(i)}$ the configuration that results from $\bm{\epsilon}$ after a collision at pair $(i,i+1)$ with random collision parameter $\alpha$, the local detailed balance condition reads
\be
P_\text{eq}(\bm{\epsilon}) \omega(\bm{\epsilon}_{\alpha}^{(i)} | \bm{\epsilon} ) = 
P_\text{eq}(\bm{\epsilon}_{\alpha}^{(i)}) \omega(\bm{\epsilon} | \bm{\epsilon}_{\alpha}^{(i)} ) \textrm{e}^{2 \chi q(\bm{\epsilon}_{\alpha}^{(i)} | \bm{\epsilon})} \, , 
\label{LDB}
\ee
where $P_\text{eq}(\bm{\epsilon})$ is the equilibrium microcanonical invariant measure, $\omega(\bm{\epsilon}' | \bm{\epsilon})$ is the transition rate for the $\bm{\epsilon} \to \bm{\epsilon}'$ jump, namely
\ben
\hspace{-0.5cm}&&\omega(\bm{\epsilon}' | \bm{\epsilon}) = \sum_{k=1}^L \int_0^1 d\alpha \, \pi(\nu_k) f_\chi(\alpha|\nu_k)  \times \label{transitionrate} \\
\hspace{-0.5cm}&& \prod_{j=1 \atop j\ne k, k+1}^L \delta(\epsilon'_j-\epsilon_j) \delta[\epsilon'_k-\alpha\nu_k] \delta[\epsilon'_{k+1}-(1-\alpha)\nu_k] \, ,  \nonumber
\een
and $q(\bm{\epsilon}' | \bm{\epsilon})$ is the energy current in the transition. For a given transition $\bm{\epsilon} \to \bm{\epsilon}_{\alpha}^{(i)}$, the current is
\be
q(\bm{\epsilon}_{\alpha}^{(i)} | \bm{\epsilon}) = \epsilon_{i} - \alpha \nu_{i} = (1-\alpha) \epsilon_{i} - \alpha \epsilon_{i+1} \, ,
\label{current}
\ee
with $\omega(\bm{\epsilon}_{\alpha}^{(i)} | \bm{\epsilon}) = \pi(\nu_i) f_\chi(\alpha|\nu_i)$. Defining now the pre-collisional energy fraction at site $i$ as $\beta\equiv \epsilon_i/\nu_i$, the forward collision amounts to a jump $\beta\to\alpha$, and the associated current can be written as $q(\bm{\epsilon}_{\alpha}^{(i)} | \bm{\epsilon})=\nu_i(\beta-\alpha)$, see Eq.~\eqref{current}, while the reverse jump is $\alpha\to\beta$. Since the microcanonical equilibrium measure is uniform on the constant-energy surface, and the collision kernel $\pi(\nu_i)$ is the same in the forward and reverse transitions, the local detailed balance condition~\eqref{LDB} reduces to
\be
\frac{f_\chi(\alpha|\nu_i)}{f_\chi(\beta|\nu_i)} = \textrm{e}^{2\chi\nu_i(\beta-\alpha)} \, . \nonumber
\label{LDBalphas}
\ee
This functional equation must hold $\forall \alpha,\beta\in[0,1]$, and hence $f_\chi(\alpha|\nu_i)\textrm{e}^{2\chi\nu_i\alpha}$ must be a constant independent of $\alpha$. Normalization in the unit interval then leads to 
\be
f_\chi(\alpha|\nu_i) = \frac{\displaystyle 2 \chi \nu_i} {\displaystyle 1 - \textrm{e}^{-2\chi\nu_i}} \textrm{e}^{-2\chi\nu_i \alpha} \, ,
\label{alphapdf}
\ee
compatible with the unbiased KMP limit $f_{\chi\to0}(\alpha|\nu_i)=1$ \cite{kipnis82a}. A similar expression has been used to introduce asymmetry in the standard (linear) KMP model~\cite{carinci16a}.

\section{Hydrodynamics}
\label{sec3}

The probability density $P_\tau(\bm{\epsilon})$ for finding the system in configuration $\bm{\epsilon}$ at time $\tau$ evolves according to the standard continuous-time master equation \cite{kampen11a}
\be
\partial_\tau P_\tau(\bm{\epsilon})= \int d\bm{\epsilon}' \Big[ \omega(\bm{\epsilon}|\bm{\epsilon}') P_\tau(\bm{\epsilon}') -  \omega(\bm{\epsilon}'|\bm{\epsilon}) P_\tau(\bm{\epsilon})\Big] \, ,
\label{mastereq}
\ee
with the transition rate $\omega(\bm{\epsilon}'|\bm{\epsilon})$ given in Eq.~\eqref{transitionrate}. 
and the total exit rate from a configuration defined above, $L \Omega(\bm{\epsilon})  = \int d\bm{\epsilon}'  \omega(\bm{\epsilon}'|\bm{\epsilon}) = \sum_{k=1}^L \pi(\nu_k)$. We are interested in the time evolution of the average energy at a given lattice site,
\be
\la \epsilon_i\ra_\tau \equiv \int d\bm{\epsilon} \, \epsilon_i \, P_\tau(\bm{\epsilon}) \, .
\label{averE}
\ee
Taking the time derivative and using the master equation \eqref{mastereq}, we obtain
\be
\partial_\tau \la \epsilon_i\ra_\tau = \int d\bm{\epsilon} \, \Delta_i(\bm{\epsilon}) \, P_\tau(\bm{\epsilon}) \, ,
\label{Deltak1}
\ee
where we have defined the instantaneous rate of change of the energy at site $i$ conditioned on the original configuration $\bm{\epsilon}$ as
\be
\Delta_i(\bm{\epsilon}) \equiv \int d\bm{\epsilon}' \, (\epsilon'_i-\epsilon_i) \, \omega(\bm{\epsilon}'|\bm{\epsilon}) \, .
\label{Deltak2}
\ee
The energy at site $i$ can only change if a collision occurs at pairs $(i-1,i)$ or $(i,i+1)$, see Eq.~\eqref{transitionrate}. A straightforward calculation then yields
\ben
\Delta_i(\bm{\epsilon})&=& \int_0^1 d\alpha \left[(1-\alpha)\nu_{i-1}-\epsilon_i\right] \pi(\nu_{i-1})  f_\chi(\alpha|\nu_{i-1})  \nonumber \\
&-& \int_0^1 d\alpha (\epsilon_i-\alpha\nu_i) \pi(\nu_i) f_\chi(\alpha|\nu_i) \, .
\label{Delta_i}
\een
Noting that the expressions inside brackets in each integral is nothing but the microscopic energy current exchanged during the corresponding collision, i.e. $q(\bm{\epsilon}_{\alpha}^{(i-1)} | \bm{\epsilon}) = (1-\alpha)\nu_{i-1}-\epsilon_i = \epsilon_{i-1}-\alpha\nu_{i-1}$ and $q(\bm{\epsilon}_{\alpha}^{(i)} | \bm{\epsilon}) = \epsilon_i-\alpha\nu_i$, see Eq.~\eqref{current}, we finally obtain 
\be
\partial_\tau \la \epsilon_i\ra_\tau = \la q_{i-1}\ra_\tau - \la q_i\ra_\tau \, ,
\label{continuityeq}
\ee
where we have defined the average current
\ben
\la q_i\ra_\tau &\equiv& \int d\bm{\epsilon} \, P_\tau(\bm{\epsilon}) \pi(\nu_i) \int_0^1 d\alpha f_\chi(\alpha|\nu_i) 
\left(\epsilon_i-\alpha\nu_i\right) \nonumber \\
&=& \int d\bm{\epsilon} \, P_\tau(\bm{\epsilon}) \pi(\nu_i) \left[\epsilon_i - \nu_i \bar{\alpha}_\chi(\nu_i) \right]\, ,
\label{avecurr}
\een
and similarly for $\la q_{i-1}\ra_\tau$, in terms of the average collision parameter 
\be
\bar{\alpha}_\chi(\nu_i) \equiv  \int_0^1 d\alpha \, \alpha \, f_\chi(\alpha|\nu_i) = \frac{1}{2\chi \nu_i} - \frac{1}{\textrm{e}^{2\chi\nu_i}-1}\, .
\label{avealpha}
\ee
Eq.~\eqref{continuityeq} is the continuity equation expressing the local conservation of energy. 

To continue, we must notice first that in order to have a balanced competition between gradient diffusion and drift in the hydrodynamic scaling limit (see below), we must consider \emph{weak} external fields $\chi=E/L$, decaying as the inverse of the system size $L$, with a constant $E$. In this case, for large $L$ the average collision parameter is then $\bar{\alpha}_\chi(\nu_i) \simeq 1/2 - E\nu_i/(6L)$, and the average current reads
\be
\la q_i\ra_\tau = \int d\bm{\epsilon} \, P_\tau(\bm{\epsilon}) \pi(\nu_i) \left[\frac{1}{2}(\epsilon_i-\epsilon_{i+1}) + \frac{E}{6L}\nu_i^2 \right]\, ,
\label{avecurr2}
\ee
where we recall that $\nu_i=\epsilon_i+\epsilon_{i+1}$. The first term is the usual diffusive contribution generated by local energy gradients, typically $\mathcal{O}(L^{-1})$, while the second one is the bulk drift induced by the weak asymmetric redistribution rule, also $\mathcal{O}(L^{-1})$. Importantly, the latter can sustain a current even for a spatially homogeneous profile on a ring, something that cannot be produced by purely boundary-driven dynamics.

We are interested now in the hydrodynamic scaling limit of the previous equations, obtained by defining the diffusively-scaled space and time variables, $x=i/L$ and $t=\tau/L^2$, respectively, with a spacing $\Delta x = 1/L$, and taking the large system size limit, $L\to\infty$. We expect that, in this hydrodynamic scaling limit, the average energy~\eqref{averE} will become a smooth field of the rescaled space and time variables, namely $\la \epsilon_i\ra_\tau \to \rho(x,t)$. Similarly, the average current~\eqref{avecurr2} will become a smooth field of $x$ and $t$, and the discrete gradient structure in the average, see Eq.~\eqref{avecurr2}, suggests $\la q_i\ra_\tau \to L^{-1} j(x,t)$. Moreover, this gradient structure explains the need of a weak field $\chi\propto L^{-1}$ for a balanced drift-diffusion competition: smooth hydrodynamic gradients generate microscopic currents of order $L^{-1}$, and the asymmetric bias must therefore be of the same order. If instead $\chi$ were kept finite as $L\to\infty$, the drift term would dominate on diffusive scales; the microscopic process would remain well defined, but the diffusive hydrodynamic description derived here would no longer apply. Noting that $\la q_{i-1}\ra_\tau \to L^{-1} j(x-\Delta x,t)$, we obtain that the local gradient in the continuity equation~\eqref{continuityeq} scales as $ \la q_{i-1}\ra_\tau - \la q_i\ra_\tau \to - L^{-2}\partial_x j(x,t)$, which together with $\partial_\tau \la \epsilon_i\ra_\tau \to L^{-2}\partial_t \rho(x,t)$, leads to the macroscopic balance equation \cite{de-groot13a, zarate06a}
\be
\partial_t \rho(x,t) + \partial_x j(x,t) = 0\, .
\label{balanceeq}
\ee

Our next task is to find the constitutive relation for the current in terms of the energy density field, computing along the way the associated transport coefficients. To proceed, we will make use of the local equilibrium (LE) approximation for the probabilities $P_\tau(\bm{\epsilon})$ \cite{de-groot13a, zarate06a, prados12a, gutierrez-ariza19a}. It is based on the clear separation of time scales that appears for large system sizes: microscopic collisions rapidly relax the system to a local equilibrium characterized by the instantaneous local average energies $\la\epsilon_i\ra_\tau$, while the relaxation of these averages toward the stationary profile occurs on a much longer, hydrodynamic time scale. Consequently, at times well beyond the microscopic time scale (defined by the average inverse exit rate $\la (L \Omega)^{-1}\ra_\tau$) the configuration probability can be approximated by a local equilibrium product measure at leading order. Subleading corrections of $\mathcal{O}(L^{-1})$ can be neglected when computing local averages, as they do not affect the leading hydrodynamic behavior. In our case, if $\bm{\epsilon}_{\widehat{i,i+1}}=(\epsilon_1,\ldots,\epsilon_{i-1},\epsilon_{i+2},\ldots,\epsilon_L)$ is the \emph{patched configuration} that results from removing the pair $(i,i+1)$ from configuration $\bm{\epsilon}$, we can write $P_\tau(\bm{\epsilon})=P_\tau(\epsilon_i,\epsilon_{i+1}|\bm{\epsilon}_{\widehat{i,i+1}}) P_\tau(\bm{\epsilon}_{\widehat{i,i+1}})$ using Bayes theorem. For large system sizes and under the local equilibrium approximation, the probability of a local configuration $(\epsilon_i,\epsilon_{i+1})$ will be independent of the rest of the system, being a locally Gibbsian measure controlled solely by the instantaneous averages $\la\epsilon_i\ra_\tau$ and $\la\epsilon_{i+1}\ra_\tau$, i.e. $P_\tau(\epsilon_i,\epsilon_{i+1}|\bm{\epsilon}_{\widehat{i,i+1}}) \approx P_\tau^{\textrm{(LE)}}(\epsilon_i,\epsilon_{i+1}) $ with
\be
P_\tau^{\textrm{(LE)}}(\epsilon_i,\epsilon_{i+1}) = \frac{\textrm{e}^{-\epsilon_i/\la\epsilon_i\ra_\tau}}{\la\epsilon_i\ra_\tau} \frac{\textrm{e}^{-\epsilon_{i+1}/\la\epsilon_{i+1}\ra_\tau}}{\la\epsilon_{i+1}\ra_\tau} \, .
\label{LE}
\ee
The average current~\eqref{avecurr2} can be then written as 
\ben
\la q_i\ra_\tau &=& \int_0^\infty d\epsilon_i d\epsilon_{i+1} \, P_\tau^{\textrm{(LE)}}(\epsilon_i,\epsilon_{i+1}) \pi(\epsilon_i + \epsilon_{i+1}) \times \nonumber \\ 
&&\left[\frac{1}{2}(\epsilon_i-\epsilon_{i+1}) + \frac{E}{6L}(\epsilon_i + \epsilon_{i+1})^2 \right]\, .
\label{avecurr3}
\een
Noticing that in the hydrodynamic scaling limit $\la \epsilon_{i+1}\ra_\tau \to \rho(x,t) + L^{-1}\partial_x \rho(x,t)$, we can now write the local equilibrium measure~\eqref{LE} as
\ben
P_\tau^{\textrm{(LE)}}(\epsilon_i,\epsilon_{i+1}) &\underset{L\gg 1}{\approx}& \frac{\textrm{e}^{-(\epsilon_i+\epsilon_{i+1})/\rho}}{\rho^2} \times  \\
&& \left[1 - \frac{1}{L} \frac{\partial_x \rho}{\rho} \left(1 - \frac{\epsilon_{i+1}}{\rho} \right) \right] \, . \nonumber
\een
The first term of this expansion is symmetric under the exchange $\epsilon_i \leftrightarrow \epsilon_{i+1}$, while the second term has no symmetry. This implies that the $\mathcal{O}(L^0)$ integral that appears when using this expansion in Eq.~\eqref{avecurr3} vanishes (note the antisymmetric term $(\epsilon_i-\epsilon_{i+1})/2$), and two different integrals appear at order $\mathcal{O}(L^{-1})$,
\begin{widetext}
\ben
\la q_i\ra_\tau^{\textrm{grad}} &=& - \frac{1}{L} \frac{\partial_x\rho}{2\rho^3} \int_0^\infty d\epsilon_i d\epsilon_{i+1} \pi(\epsilon_i + \epsilon_{i+1})\left(1- \frac{\epsilon_{i+1}}{\rho} \right) (\epsilon_i-\epsilon_{i+1}) \textrm{e}^{-(\epsilon_i+\epsilon_{i+1})/\rho}  \nonumber \\ 
&=& - \frac{1}{L} \frac{\partial_x\rho}{4\rho^4} \int_0^\infty d\epsilon_i d\epsilon_{i+1} \pi(\epsilon_i + \epsilon_{i+1}) (\epsilon_i-\epsilon_{i+1})^2 \textrm{e}^{-(\epsilon_i+\epsilon_{i+1})/\rho}  \, ,\label{avecurr4a} \\
\la q_i\ra_\tau^{\textrm{drift}} &=& + \frac{1}{L} \frac{E}{6\rho^2} \int_0^\infty d\epsilon_i d\epsilon_{i+1} \pi(\epsilon_i + \epsilon_{i+1}) (\epsilon_i+\epsilon_{i+1})^2 \textrm{e}^{-(\epsilon_i+\epsilon_{i+1})/\rho}  \, ,
\label{avecurr4b}
\een
which are the gradient and drift contributions to the current. Note that the second line in Eq.~\eqref{avecurr4a} corresponds to the symmetrized version of the first line, as this first integral remains invariant under the exchange of the dummy variables $\epsilon_i \leftrightarrow \epsilon_{i+1}$. The two integrals~\eqref{avecurr4a}-\eqref{avecurr4b} can be now computed by two consecutive changes of variables, first $\epsilon_i=\rho\, a$ and $\epsilon_{i+1}=\rho\, b$, and then $\sqrt{a}=r\cos\phi$ and $\sqrt{b}=r\sin\phi$, with polar coordinates in the domain $r\in[0,\infty)$ and $\phi\in[0,\pi/2]$. This leads to
\ben
\la q_i\ra_\tau^{\textrm{grad}} &=& - \frac{\partial_x\rho}{L} \underbrace{\int_0^{\frac{\pi}{2}} d\phi \sin\phi\cos\phi \cos^2(2\phi)}_{1/6} \int_0^\infty dr r^7  \pi(\rho r^2) \textrm{e}^{-r^2} = - \frac{\partial_x\rho}{L} \left( \frac{1}{6} \int_0^\infty dr r^7  \pi(\rho r^2) \textrm{e}^{-r^2}\right) ,\label{avecurr5a} \\
\la q_i\ra_\tau^{\textrm{drift}} &=& + \frac{E}{L} \frac{2\rho^2}{3} \underbrace{\int_0^{\frac{\pi}{2}} d\phi \sin\phi\cos\phi}_{1/2} \int_0^\infty dr r^7  \pi(\rho r^2) \textrm{e}^{-r^2} = + \frac{E}{L} \left( \frac{\rho^2}{3} \int_0^\infty dr r^7  \pi(\rho r^2) \textrm{e}^{-r^2}\right) \, .
\label{avecurr5b}
\een
\end{widetext}
In this way we confirm the hydrodynamic scaling $\la q_i\ra_\tau \to L^{-1} j(x,t)$ suggested before, with a constitutive relation for the macroscopic current
\be
j(x,t) = - D(\rho) \partial_x\rho(x,t) + \sigma(\rho) E \, ,
\label{jcurrent}
\ee
where we have defined the diffusivity $D(\rho)$ and mobility $\sigma(\rho)$ transport coefficients as
\ben
D(\rho) &=& \frac{1}{6} \int_0^\infty dr r^7  \pi(\rho r^2) \textrm{e}^{-r^2} \, , \label{diff} \\
\sigma(\rho) &=& \frac{\rho^2}{3} \int_0^\infty dr r^7  \pi(\rho r^2) \textrm{e}^{-r^2} \, . \label{mobi}
\een
These coefficients, which depend nontrivially on the function $\pi(\nu)$ controlling the microscopic collision rate depending on the pair energy, are related by an Einstein-like relation $\sigma(\rho) = 2\rho^2 D(\rho)$, as expected \cite{de-groot13a, zarate06a, prados12a, gutierrez-ariza19a}. Moreover the mobility, which modulates here the system response to a weak external field, also controls the fluctuations of the local current field, as already demonstrated for the nonlinear but symmetric version of the KMP model \cite{hurtado12a, prados12a}.

\section{Taming nonlinear energy diffusion}

Our next objective consists in validating the driving mechanism in simulations, testing in detail the response of the nonlinear energy diffusion model to an external field as the one described in Section \S\ref{sec3}. Furthermore, as a proof of concept, we will use this driving mechanism to create a \emph{packing field} around the instantaneous center of mass of the energy field in the $1d$ ring. It can be shown that, for sufficiently large couplings, this packing field triggers a nonequilibrium phase transition to a programmable time-crystal phase \cite{hurtado-gutierrez20a, hurtado-gutierrez23a, hurtado-gutierrez25a, hurtado-gutierrez25b, hurtado25a}, characterized by the emergence of traveling condensates that break continuous time-translation symmetry. We will show here how such nonequilibrium phases also appear for this bulk-driven nonlinear energy diffusion model.

\subsection{Constant-field driving}

We first consider the response of the system to a constant external field $E$. For periodic boundary conditions, the steady state under this constant driving will be homogeneous, $\la \rho\ra_\text{st}=\rho_0$, with $\rho_0$ the average energy density. In the absence of spatial gradients, the stationary average current will be thus simply $\la j\ra_\text{st} = \sigma(\rho_0) E$, so measuring the system response to an external field for different average densities $\rho_0$ will allow us to obtain the mobility transport coefficient,
\be
\sigma(\rho_0) = \frac{\la j\ra_\text{st}}{E} \, .
\label{mobiMC}
\ee

We hence performed extensive kinetic Monte Carlo simulations of the driven nonlinear energy diffusion model for different constant fields $E$, system sizes $L$ and global  densities $\rho_0$, measuring $\la j\ra_\text{st}$ to obtain $\sigma(\rho_0)$ and corroborate our prediction~\eqref{mobi} (the diffusivity coefficient was already validated in boundary driving simulations, see e.g.~\cite{prados12a}). To be concrete, hereafter we restrict our results to a broad class of energy diffusion models characterized by a power-law kernel controlling the microscopic collision rate \cite{prados11a, prados12a, hurtado13a}
\be
\pi(\nu) = \frac{2}{\Gamma(\beta+3)}\nu^\beta \, .
\label{pifunc}
\ee
with an exponent $\beta>-3$ but otherwise arbitrary \cite{prados12a}. This restriction is related to the convergence of the integrals defining the transport coefficients~\footnote{In particular, while the convergence of the integrals~\eqref{diff}-\eqref{mobi} for the diffusivity and mobility transport coefficients in terms of the microscopic collision kernel $\pi(\nu)$ demands $\beta>-4$, the convergence of the energy-dissipation coefficient for the granular or dissipative generalization of the nonlinear KMP model impose the stiffer restriction $\beta>-3$, see Ref. \cite{prados12a}.}. Note that for $\beta=0$ we recover the constant, energy-independent collision rate of the original (linear) KMP model \cite{kipnis82a}. For this particular collision kernel~\eqref{pifunc} the diffusivity and mobility are simply
\be
D(\rho) = \frac{\beta+3}{6}\rho^\beta \, , \quad \sigma(\rho) = \frac{\beta+3}{3}\rho^{\beta+2} \, ,
\label{coeffs}
\ee
see Eqs.~\eqref{diff}-\eqref{mobi}.

\begin{figure}[t]
\includegraphics[width=8cm]{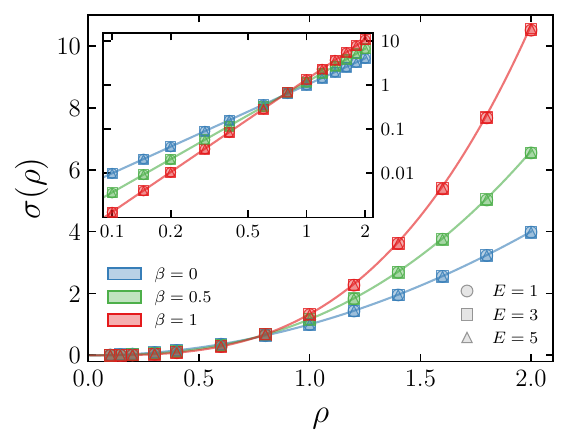}
\vspace{-0.75cm}
\caption{Main panel: Mobility transport coefficient $\sigma(\rho)$ as a function of the density, measured in kinetic Monte Carlo simulations of the driven nonlinear energy diffusion model on a periodic $1d$ lattice with $L=200$ sites. Different symbols correspond to external fields $E=1$ ($\bigcirc$), $E=3$ ($\Box$) and $E=5$ ($\triangle$), while different colors indicate the nonlinearity exponent $\beta=0$, $0.5$ and $1$. Inset: the same data in log-log scale, to highlight the power law behavior $\sigma(\rho)\sim\rho^{\beta+2}$.}
\label{fig1}
\end{figure}

Fig.~\ref{fig1} shows the results obtained for the mobility~\eqref{mobiMC} for a lattice size $L=200$, three different exponent values $\beta=0,~0.5,~1$, varying global densities, and different values of the constant driving field $E=1,~3,~5$. These values were chosen within the weak-field regime, i.e. $|E \nu/L| \ll 1$ for the typical pair energy $\nu$, see Eq.~\eqref{avecurr2}, where the diffusive hydrodynamic prediction is expected to hold quantitatively. Other values of $L$ were simulated but no appreciable finite-size effects were observed. The agreement with the theoretical predictions is excellent in all cases, with a mobility coefficient that increases as a power law with the energy density, $\sigma(\rho)\sim\rho^{\beta+2}$, as expected (see inset to Fig.~\ref{fig1}). This confirms the hydrodynamic drift derived within the local equilibrium approximation from a bulk driving mechanism based on a tilted collision parameter distribution, see Eq.~\eqref{alphapdf}.

\subsection{Packing driving field and time crystals}

Our next task consists in demonstrating how this bulk driving mechanism opens up new possibilities to control nonlinear energy diffusion, in particular by creating nonequilibrium time crystal phases. Broadly speaking, time crystals are many-body systems that break spontaneously time-translation symmetry \cite{wilczek12a, zakrzewski12a, sacha18a, sacha20a}, resulting in robust long-range temporal order and persistent ground-state periodic motion. Recent work on rare current fluctuations in many-body systems \cite{bodineau05a, hurtado11a, perez-espigares13a, hurtado14a, perez-espigares19a, hurtado25a} has shown that a so-called \emph{packing field} selectively amplifying density fluctuations can induce time-crystalline spatiotemporal order in driven diffusive media \cite{hurtado-gutierrez20a, hurtado-gutierrez23a, hurtado-gutierrez25a, hurtado-gutierrez25b}. We now explore this phenomenon in the driven nonlinear energy diffusion model studied in this paper.

More in detail, we consider now an external field with two components, $E_x[\rho]=E_0 + \lambda \mathcal{E}_x^{(m)}[\rho]$. The first component is a constant bulk field $E_0$ driving homogeneously energy flow in a given direction, while $\lambda\ge 0$ is the coupling constant to a spatially-structured $m$th-order packing field 
\be
\mathcal{E}_x^{(m)}[\rho] = |z_m| \sin \left(\varphi_m - 2\pi m x \right) \, .
\label{packingfield}
\ee
In this expression $z_m$ is the complex $m$th-order packing order parameter for a given field state $\rho(x,t)$,
\be
z_m[\rho] = \rho_0^{-1} \int_0^1 dx \, \rho(x,t) \, \textrm{e}^{\textrm{i} 2\pi m x} \equiv |z_m| \textrm{e}^{\textrm{i}\varphi_m} \, ,
\label{packingOPd}
\ee
of magnitude $|z_m|$ and argument $\varphi_m$. The working of the packing field~\eqref{packingfield} is more easily understood for $m=1$. In this case $\mathcal{E}_x^{(1)}[\rho]$ is a single-mode sine function pushing energy that lags behind the instantaneous center of mass of the energy field $\rho(x,t)$, located at $\varphi_1$, while restraining that moving ahead. This leads to a nonlinear feedback loop that amplifies naturally-occurring fluctuations of the energy field's spatial packing, as measured by $|z_1|$, eventually triggering a transition to a time crystal phase characterized by the emergence of a traveling energy condensate. From a mathematical perspective, $\mathcal{E}_x^{(1)}[\rho]$ can be understood as a controlled excitation of the first Fourier mode of the energy field around the center-of-mass position, so it is expected that an equivalent excitation of higher, $m$th-order modes ($m>1$) via $\mathcal{E}_x^{(m)}[\rho]$ will generate fully programmable continuous time-crystal phases in driven diffusive media \cite{hurtado-gutierrez25a, hurtado-gutierrez25b} with $m$ emergent traveling energy condensates. This idea has been confirmed for a broad range of driven diffusive models using hydrodynamics tools \cite{hurtado-gutierrez25a, hurtado-gutierrez25b}. Indeed, analyzing the linear stability of the homogeneous density solution under $m$th-order packing fields leads to a critical coupling $\lambda_c^{(m)}$ beyond which the time crystal transition occurs \cite{hurtado-gutierrez20a,hurtado-gutierrez25a},
\be
\lambda_c^{(m)} = 4\pi m \frac{D(\rho_0)\rho_0}{\sigma(\rho_0)} = 2\pi m \rho_0^{-1}\, ,
\label{criticalpoint}
\ee
where we have used the explicit transport coefficients~\eqref{coeffs} in the second equality. 

\begin{figure}[t]
\includegraphics[width=8.1cm]{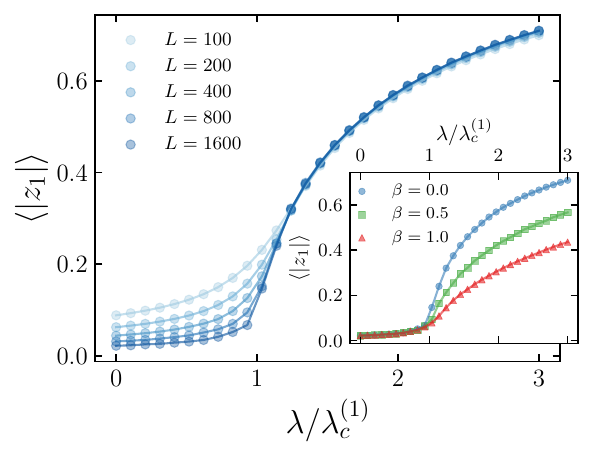}
\vspace{-0.75cm}
\caption{Main: Packing order parameter $\la |z_1|\ra$ as a function of the reduced coupling $\lambda/\lambda_c^{(1)}$ measured in simulations for $\beta=0$, $\rho_0=1/3$, $E_0=10$ and different values of $L$. Inset: $\la |z_1|\ra$ vs $\lambda/\lambda_c^{(1)}$ measured for $L=1600$, varying $\beta=0,~0.5,~1.0$ and the same $\rho_0=1/3$ and $E_0=10$. Error bars smaller than symbol sizes in all cases.}
\label{fig2}
\end{figure}

\begin{figure}
\vspace{-0.1cm}
\includegraphics[width=8.1cm,clip]{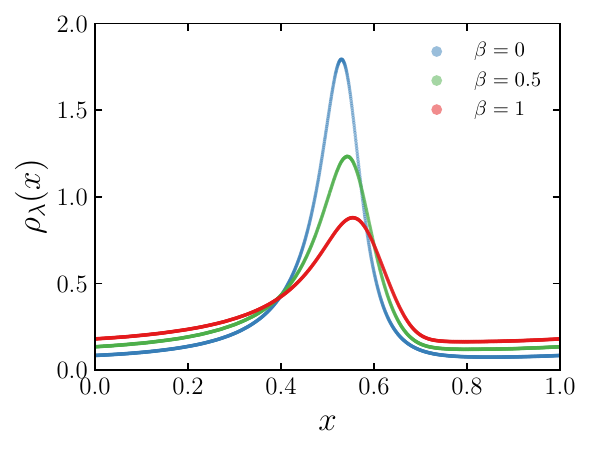}
\includegraphics[width=8.1cm,clip,trim=0 1cm 0 2cm]{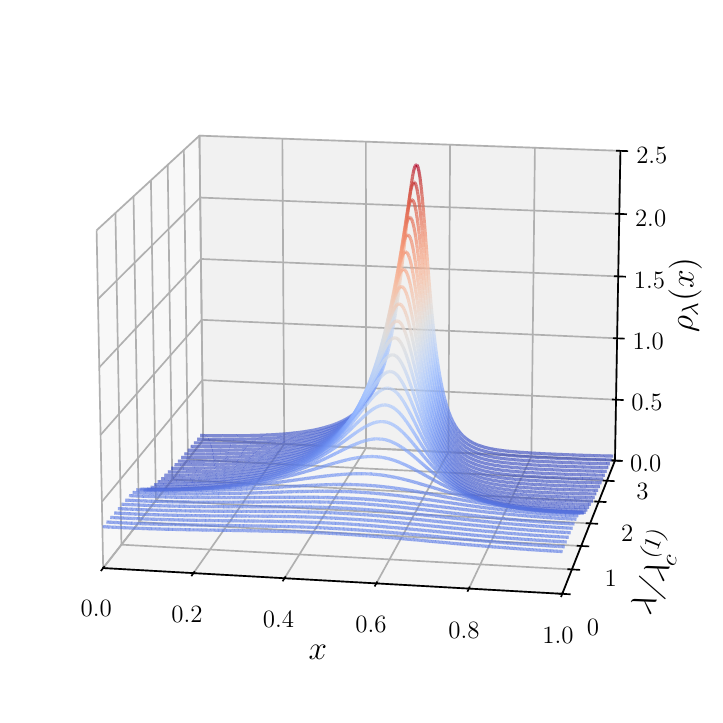}
\caption{Top panel: Average energy density field $\rho_\lambda(x)$ around the center of mass for supercritical coupling $\lambda/\lambda_c^{(1)}=2.379$ and different nonlinearity exponents $\beta=0,~0.5,~1.0$, as measured in simulations for $L=1600$, $\rho_0=1/3$ and $E_0=10$. Error bars are smaller than symbol sizes. Bottom panel: Evolution of the average energy density field (centered around the instantaneous center of mass) with the reduced coupling $\lambda/\lambda_c^{(1)}$ measured for $\beta=0$, $L=800$, $\rho_0=1/3$ and $E_0=10$.}
\label{fig3}
\end{figure}

We now study this phenomenon microscopically by means of extensive kinetic Monte Carlo simulations of the nonlinear energy diffusion model driven by an $m$th-order packing field. Figure~\ref{fig2} provides the first evidence for the formation of a time-crystal phase under a first-order packing field ($m=1$). The main panel shows the average magnitude of the packing order parameter $\la |z_1|\ra$ as a function of the reduced coupling $\lambda/\lambda_c^{(1)}$ measured for $\beta=0$, average density $\rho_0=1/3$, constant bulk field $E_0=10$ and different system sizes. This order parameter, that measures the level of packing of the energy field around its instantaneous center of mass, vanishes below the critical coupling and grows continuously beyond $\lambda_c^{(1)}$, displaying clear signatures of a second-order phase transition. Finite-size effects are progressively suppressed as $L$ increases, with curves converging toward a well-defined thermodynamic limit, consistent with the hydrodynamic scenario \cite{hurtado-gutierrez25a} and confirming the predicted critical threshold $\lambda_c^{(1)}$, see Eq.~\eqref{criticalpoint}. The inset of Fig.~\ref{fig2} illustrates the dependence of the order parameter on the nonlinearity exponent $\beta$ for the largest system size measured. Interestingly, increasing $\beta$ leads to a slower growth of $\la |z_1|\ra(\lambda)$ above criticality, indicating that stronger energy-dependent collision rates soften the response to the packing field due to the enhanced role of nonlinear diffusion in spreading energy away from the packet core, see below. In any case, these results support the formation of an energy condensate packing energy locally for couplings beyond the critical one.

Further insight into the physics of the time crystal phase can be gained by studying the structure of the emerging energy condensate, see Fig.~\ref{fig3}. The top panel shows the average energy density profile around the center of mass measured for a large $L$ and a supercritical $\lambda/\lambda_c^{(1)}>1$, for the same three $\beta$ values and same parameters considered in Fig.~\ref{fig2}. The energy condensate exhibits in all cases a strongly asymmetric shape, characterized by a sharp leading front in the direction of propagation and a broader trailing tail, as predicted by hydrodynamics \cite{hurtado-gutierrez25a}. As $\beta$ increases, the condensate profile becomes progressively smoother, in agreement with the softening of the order parameter with increasing $\beta$, see inset of Fig.~\ref{fig2}. The physical reason is that increasing $\beta$ makes collisions more frequent in high-energy regions, enhancing the effective transport activity around the condensate core, which in turn facilitates the redistribution of energy away from sharply localized structures. In this way, the same nonlinearity that increases the mobility of high-energy regions also makes them less able to sustain sharp packing under the same feedback field. The bottom panel of Fig.~\ref{fig3} displays the full evolution of the condensate profile with the relative coupling $\lambda/\lambda_c^{(1)}$ measured for $\beta=0$. As expected, the critical point signals the emergence of localized structure. 

\begin{figure}
\includegraphics[width=8.5cm]{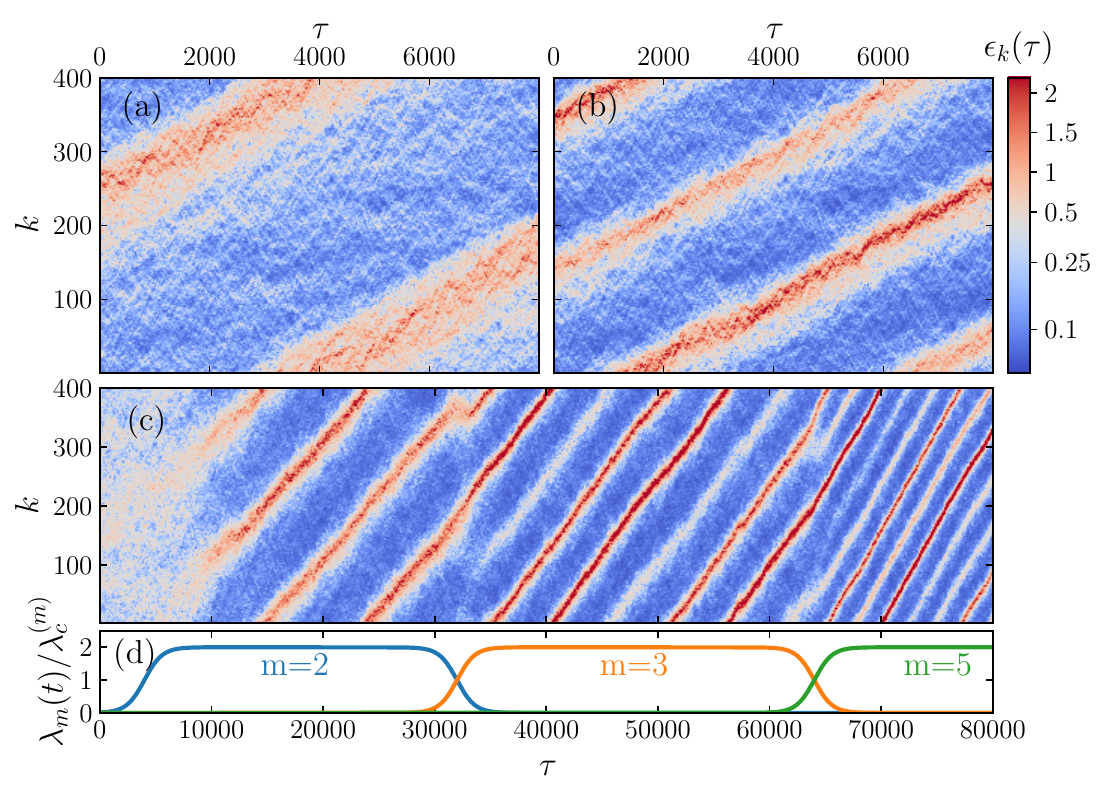}
\vspace{-0.75cm}
\caption{Spatiotemporal raster plots illustrating the dynamics of the energy field for the nonlinear energy diffusion model driven by different packing fields. All panels correspond to steady-state trajectories for $\rho_0=1/3$, $\beta=0$ and $L=400$. Panels (a) and (b) show the system response to $m=1,~2$ packing fields, respectively, with reduced coupling $\lambda/\lambda_c^{(m)}=2$ for constant field $E_0=20$. Panel (c) displays the system response to a dynamic superposition of different packing fields, see Eq.~\eqref{Emodul}, with reduced amplitudes modulated in time as shown in panel (d), for $E_0=10$. 
}
\label{fig4}
\end{figure}

This structure is also apparent at the single trajectory level, not only on average behavior. This is illustrated in Figs.~\ref{fig4}.a-b, which show two different spatiotemporal raster plots displaying typical steady state trajectories of the energy density field for packing fields of order $m=1$ and $m=2$, respectively. Note that despite the underlying stochastic dynamics, the motion of the emerging energy waves is ballistic. The packing field thus opens up the possibility of controlling the spatiotemporal response of the energy diffusion model: the constant field $E_0$ controls  the velocity of the emergent wave pattern, the integer index $m$ controls the number of emergent condensates, and the reduced coupling $\lambda/\lambda_c^{(m)}$ tunes the sharpness of energy packets. We can also force the competition of different packing fields by modulating in time their coupling constants to engineer custom and programmable time-crystal phases \cite{hurtado-gutierrez25a}, e.g. we may consider fields
\be
E_x[\rho] = E_0 + \sum_m \lambda_m(t) \mathcal{E}_x^{(m)}[\rho] \, .
\label{Emodul}
\ee
The result of the dynamic superposition of multiple packing fields is illustrated in the raster plot of Fig.~\ref{fig4}.c, where packing fields of order $m=2,~3$ and $5$ are modulated in time as shown in Fig.~\ref{fig4}.d. The ensuing energy field dynamics exhibits a rich, highly structured spatiotemporal pattern, combining multiple traveling modes as dictated by the external forcing, which enables custom programmability. These results highlight the effectiveness of packing fields as a tool to tame nonlinear energy diffusion and engineer complex nonequilibrium time-crystalline phases in stochastic energy transport models.

\section{Conclusion}

We have studied in this work a bulk-driven nonlinear generalization of the Kipnis–Marchioro–Presutti model of energy diffusion, enabling direct control over energy transport at the microscopic level. By biasing the local stochastic collision dynamics, we obtain a driven diffusive dynamics that induces a finite energy current in the bulk, mimicking the action of an external field. We have derived the hydrodynamic description of the model from the exact master equation using a local equilibrium approximation, obtaining the explicit constitutive relation for the energy current and closed expressions for the associated diffusivity and mobility coefficients, which depend nonlinearly on the local energy density via the microscopic collision rate. Extensive kinetic Monte Carlo simulations confirm the predictions, validating the hydrodynamic framework beyond linear response.

We have further demonstrated the versatility of this approach by using it to engineer interesting driven nonequilibrium phases. In particular, we have shown that suitably designed packing fields coupled to energy field fluctuations induce a continuous phase transition to a time-crystalline regime characterized by the emergence of robust traveling energy condensates. These condensates exhibit long-range spatiotemporal order and persistent periodic motion, breaking continuous time-translation symmetry despite the stochastic nature of the underlying dynamics. Importantly, we have shown how the properties of these phases—such as the number, shape, and velocity of the condensates—can be tuned through the structure of the external driving, highlighting the role of bulk fields as powerful control knobs for nonlinear energy transport.

Our results open several promising directions for future research. For instance, bulk driving enables the systematic engineering of dynamical instabilities in energy transport models, other than the time-crystal phases here studied. These ideas can be then easily extended to bulk-driven KMP-type models defined in higher dimensions. The bulk-driving framework can be generalized also to models with multiple conserved quantities, as e.g. the kinetic exclusion process \cite{gutierrez-ariza19a}, enabling the possibility of studying differential particle and energy driving. Finally, generic bulk driving offers a possible physical realization of the effective dynamics that makes typical a rare event via the abstract Doob transform \cite{doob57a, chetrite15a, chetrite15b, bertini15a, carollo18b, simon09a, jack10a, popkov10a}. Exploring this connection further could clarify the physics of stochastic processes conditioned on large deviations \cite{touchette09a, hurtado14a, bertini15a, derrida07a, perez-espigares19a, hurtado25a}, and suggest new ways to experimentally or numerically access rare-event physics.

\begin{acknowledgments}
The research leading to these results has received funding from the I+D+i grants PID2023-149365NB-I00, PID2020-113681GB-I00, and C-EXP-251-UGR23, funded by MICIU/AEI/10.13039/501100011033/, ERDF/EU, and Junta de Andaluc\'{\i}a - Consejer\'{\i}a de Econom\'{\i}a y Conocimiento. We are also grateful for the the computing resources and technical support provided by PROTEUS, the supercomputing center of the Institute Carlos I for Theoretical and Computational Physics in Granada, Spain.
\end{acknowledgments}

\bibliography{/Users/phurtado/PAPERS/BIBLIOGRAPHY/referencias-BibDesk-OK.bib}{}

\begin{thebibliography}{81}%
\makeatletter
\providecommand \@ifxundefined [1]{%
 \@ifx{#1\undefined}
}%
\providecommand \@ifnum [1]{%
 \ifnum #1\expandafter \@firstoftwo
 \else \expandafter \@secondoftwo
 \fi
}%
\providecommand \@ifx [1]{%
 \ifx #1\expandafter \@firstoftwo
 \else \expandafter \@secondoftwo
 \fi
}%
\providecommand \natexlab [1]{#1}%
\providecommand \enquote  [1]{``#1''}%
\providecommand \bibnamefont  [1]{#1}%
\providecommand \bibfnamefont [1]{#1}%
\providecommand \citenamefont [1]{#1}%
\providecommand \href@noop [0]{\@secondoftwo}%
\providecommand \href [0]{\begingroup \@sanitize@url \@href}%
\providecommand \@href[1]{\@@startlink{#1}\@@href}%
\providecommand \@@href[1]{\endgroup#1\@@endlink}%
\providecommand \@sanitize@url [0]{\catcode `\\12\catcode `\$12\catcode
  `\&12\catcode `\#12\catcode `\^12\catcode `\_12\catcode `\%12\relax}%
\providecommand \@@startlink[1]{}%
\providecommand \@@endlink[0]{}%
\providecommand \url  [0]{\begingroup\@sanitize@url \@url }%
\providecommand \@url [1]{\endgroup\@href {#1}{\urlprefix }}%
\providecommand \urlprefix  [0]{URL }%
\providecommand \Eprint [0]{\href }%
\providecommand \doibase [0]{http://dx.doi.org/}%
\providecommand \selectlanguage [0]{\@gobble}%
\providecommand \bibinfo  [0]{\@secondoftwo}%
\providecommand \bibfield  [0]{\@secondoftwo}%
\providecommand \translation [1]{[#1]}%
\providecommand \BibitemOpen [0]{}%
\providecommand \bibitemStop [0]{}%
\providecommand \bibitemNoStop [0]{.\EOS\space}%
\providecommand \EOS [0]{\spacefactor3000\relax}%
\providecommand \BibitemShut  [1]{\csname bibitem#1\endcsname}%
\let\auto@bib@innerbib\@empty
\bibitem [{\citenamefont {Anderson}(1972)}]{anderson72a}%
  \BibitemOpen
  \bibfield  {author} {\bibinfo {author} {\bibfnamefont {P.~W.}\ \bibnamefont
  {Anderson}},\ }\bibfield  {title} {\enquote {\bibinfo {title} {More is
  different},}\ }\href {\doibase 10.1126/science.177.4047.393} {\bibfield
  {journal} {\bibinfo  {journal} {Science}\ }\textbf {\bibinfo {volume}
  {177}},\ \bibinfo {pages} {393} (\bibinfo {year} {1972})}\BibitemShut
  {NoStop}%
\bibitem [{\citenamefont {Peierls}(1980)}]{peierls80a}%
  \BibitemOpen
  \bibfield  {author} {\bibinfo {author} {\bibfnamefont {R.}~\bibnamefont
  {Peierls}},\ }\bibfield  {title} {\enquote {\bibinfo {title} {Model-making in
  physics},}\ }\href {\doibase 10.1080/00107518008210938} {\bibfield  {journal}
  {\bibinfo  {journal} {Contemporary Physics}\ }\textbf {\bibinfo {volume}
  {21}},\ \bibinfo {pages} {3} (\bibinfo {year} {1980})}\BibitemShut {NoStop}%
\bibitem [{\citenamefont {Goldenfeld}\ and\ \citenamefont
  {Kadanoff}(1999)}]{goldenfeld99a}%
  \BibitemOpen
  \bibfield  {author} {\bibinfo {author} {\bibfnamefont {N.}~\bibnamefont
  {Goldenfeld}}\ and\ \bibinfo {author} {\bibfnamefont {L.~P.}\ \bibnamefont
  {Kadanoff}},\ }\bibfield  {title} {\enquote {\bibinfo {title} {Simple lessons
  from complexity},}\ }\href {\doibase 10.1126/science.284.5411.87} {\bibfield
  {journal} {\bibinfo  {journal} {Science}\ }\textbf {\bibinfo {volume}
  {284}},\ \bibinfo {pages} {87} (\bibinfo {year} {1999})}\BibitemShut
  {NoStop}%
\bibitem [{\citenamefont {Widom}(1974)}]{widom74a}%
  \BibitemOpen
  \bibfield  {author} {\bibinfo {author} {\bibfnamefont {B.}~\bibnamefont
  {Widom}},\ }\bibfield  {title} {\enquote {\bibinfo {title} {What is a
  model?}}\ }\href {\doibase 10.1126/science.183.4122.305} {\bibfield
  {journal} {\bibinfo  {journal} {Science}\ }\textbf {\bibinfo {volume}
  {183}},\ \bibinfo {pages} {305} (\bibinfo {year} {1974})}\BibitemShut
  {NoStop}%
\bibitem [{\citenamefont {Wigner}(1960)}]{wigner60a}%
  \BibitemOpen
  \bibfield  {author} {\bibinfo {author} {\bibfnamefont {E.~P.}\ \bibnamefont
  {Wigner}},\ }\bibfield  {title} {\enquote {\bibinfo {title} {The unreasonable
  effectiveness of mathematics in the natural sciences},}\ }\href {\doibase
  10.1002/cpa.3160130102} {\bibfield  {journal} {\bibinfo  {journal} {Comm.
  Pure Appl. Math.}\ }\textbf {\bibinfo {volume} {13}},\ \bibinfo {pages} {1}
  (\bibinfo {year} {1960})}\BibitemShut {NoStop}%
\bibitem [{\citenamefont {Lebowitz}(1999)}]{lebowitz99b}%
  \BibitemOpen
  \bibfield  {author} {\bibinfo {author} {\bibfnamefont {J.~L.}\ \bibnamefont
  {Lebowitz}},\ }\bibfield  {title} {\enquote {\bibinfo {title} {Microscopic
  origins of irreversible macroscopic behavior},}\ }\href {\doibase
  10.1016/S0378-4371(98)00517-7} {\bibfield  {journal} {\bibinfo  {journal}
  {Physica A:}\ }\textbf {\bibinfo {volume} {263}},\ \bibinfo {pages} {516}
  (\bibinfo {year} {1999})}\BibitemShut {NoStop}%
\bibitem [{\citenamefont {Baxter}(2016)}]{baxter16a}%
  \BibitemOpen
  \bibfield  {author} {\bibinfo {author} {\bibfnamefont {R.~J.}\ \bibnamefont
  {Baxter}},\ }\href@noop {} {\emph {\bibinfo {title} {Exactly solved models in
  statistical mechanics}}}\ (\bibinfo  {publisher} {Elsevier},\ \bibinfo {year}
  {2016})\BibitemShut {NoStop}%
\bibitem [{\citenamefont {Marro}\ and\ \citenamefont
  {Dickman}(2005)}]{marro05a}%
  \BibitemOpen
  \bibfield  {author} {\bibinfo {author} {\bibfnamefont {J.}~\bibnamefont
  {Marro}}\ and\ \bibinfo {author} {\bibfnamefont {R.}~\bibnamefont
  {Dickman}},\ }\href@noop {} {\emph {\bibinfo {title} {Nonequilibrium phase
  transitions in lattice models}}}\ (\bibinfo  {publisher} {Cambridge
  University Press},\ \bibinfo {year} {2005})\BibitemShut {NoStop}%
\bibitem [{\citenamefont {Kipnis}\ \emph {et~al.}(1982)\citenamefont {Kipnis},
  \citenamefont {Marchioro},\ and\ \citenamefont {Presutti}}]{kipnis82a}%
  \BibitemOpen
  \bibfield  {author} {\bibinfo {author} {\bibfnamefont {C.}~\bibnamefont
  {Kipnis}}, \bibinfo {author} {\bibfnamefont {C.}~\bibnamefont {Marchioro}}, \
  and\ \bibinfo {author} {\bibfnamefont {E.}~\bibnamefont {Presutti}},\
  }\bibfield  {title} {\enquote {\bibinfo {title} {Heat-flow in an exactly
  solvable model},}\ }\href
  {http://link.springer.com/article/10.1007/BF01011740} {\bibfield  {journal}
  {\bibinfo  {journal} {J. Stat. Phys.}\ }\textbf {\bibinfo {volume} {27}},\
  \bibinfo {pages} {65--74} (\bibinfo {year} {1982})}\BibitemShut {NoStop}%
\bibitem [{\citenamefont {Bonetto}\ \emph {et~al.}(2000)\citenamefont
  {Bonetto}, \citenamefont {Lebowitz},\ and\ \citenamefont
  {Rey-Bellet}}]{bonetto00a}%
  \BibitemOpen
  \bibfield  {author} {\bibinfo {author} {\bibfnamefont {F.}~\bibnamefont
  {Bonetto}}, \bibinfo {author} {\bibfnamefont {J.~L.}\ \bibnamefont
  {Lebowitz}}, \ and\ \bibinfo {author} {\bibfnamefont {L.}~\bibnamefont
  {Rey-Bellet}},\ }\enquote {\bibinfo {title} {Mathematical physics 2000},}\ \
  (\bibinfo  {publisher} {Imperial College Press},\ \bibinfo {address}
  {London},\ \bibinfo {year} {2000})\ Chap.\ \bibinfo {chapter} {{F}ourier's
  law: {A} challenge for theorists}, p.\ \bibinfo {pages} {128}\BibitemShut
  {NoStop}%
\bibitem [{\citenamefont {Lepri}\ \emph {et~al.}(2003)\citenamefont {Lepri},
  \citenamefont {Livi},\ and\ \citenamefont {Politi}}]{lepri03a}%
  \BibitemOpen
  \bibfield  {author} {\bibinfo {author} {\bibfnamefont {S.}~\bibnamefont
  {Lepri}}, \bibinfo {author} {\bibfnamefont {R.}~\bibnamefont {Livi}}, \ and\
  \bibinfo {author} {\bibfnamefont {A.}~\bibnamefont {Politi}},\ }\bibfield
  {title} {\enquote {\bibinfo {title} {Thermal conduction in classical
  low-dimensional lattices},}\ }\href {\doibase 10.1016/S0370-1573(02)00558-6}
  {\bibfield  {journal} {\bibinfo  {journal} {Phys. Rep.}\ }\textbf {\bibinfo
  {volume} {377}},\ \bibinfo {pages} {1--80} (\bibinfo {year}
  {2003})}\BibitemShut {NoStop}%
\bibitem [{\citenamefont {Dhar}(2008)}]{dhar08a}%
  \BibitemOpen
  \bibfield  {author} {\bibinfo {author} {\bibfnamefont {A.}~\bibnamefont
  {Dhar}},\ }\bibfield  {title} {\enquote {\bibinfo {title} {Heat transport in
  low-dimensional systems},}\ }\href
  {http://www.tandfonline.com/doi/abs/10.1080/00018730802538522} {\bibfield
  {journal} {\bibinfo  {journal} {Adv. Phys.}\ }\textbf {\bibinfo {volume}
  {57}},\ \bibinfo {pages} {457} (\bibinfo {year} {2008})}\BibitemShut
  {NoStop}%
\bibitem [{\citenamefont {Lepri}(2016)}]{lepri16a}%
  \BibitemOpen
  \bibinfo {editor} {\bibfnamefont {Stefano}\ \bibnamefont {Lepri}},\ ed.,\
  \href {\doibase 10.1007/978-3-319-29261-8} {\emph {\bibinfo {title} {Thermal
  Transport in Low Dimensions: From Statistical Physics to Nanoscale Heat
  Transfer}}},\ \bibinfo {series} {Lectures Notes in Physics}, Vol.\ \bibinfo
  {volume} {921}\ (\bibinfo  {publisher} {Springer},\ \bibinfo {year}
  {2016})\BibitemShut {NoStop}%
\bibitem [{\citenamefont {Derrida}\ and\ \citenamefont
  {Lebowitz}(1998)}]{derrida98b}%
  \BibitemOpen
  \bibfield  {author} {\bibinfo {author} {\bibfnamefont {B.}~\bibnamefont
  {Derrida}}\ and\ \bibinfo {author} {\bibfnamefont {J.~L.}\ \bibnamefont
  {Lebowitz}},\ }\bibfield  {title} {\enquote {\bibinfo {title} {Exact large
  deviation function in the asymmetric exclusion process},}\ }\href@noop {}
  {\bibfield  {journal} {\bibinfo  {journal} {Phys. Rev. Lett.}\ }\textbf
  {\bibinfo {volume} {80}},\ \bibinfo {pages} {209--213} (\bibinfo {year}
  {1998})}\BibitemShut {NoStop}%
\bibitem [{\citenamefont {Derrida}\ \emph {et~al.}(2001)\citenamefont
  {Derrida}, \citenamefont {Lebowitz},\ and\ \citenamefont
  {Speer}}]{derrida01a}%
  \BibitemOpen
  \bibfield  {author} {\bibinfo {author} {\bibfnamefont {B.}~\bibnamefont
  {Derrida}}, \bibinfo {author} {\bibfnamefont {J.~L.}\ \bibnamefont
  {Lebowitz}}, \ and\ \bibinfo {author} {\bibfnamefont {E.~R.}\ \bibnamefont
  {Speer}},\ }\bibfield  {title} {\enquote {\bibinfo {title} {Free energy
  functional for nonequilibrium systems: {An} exactly solvable case},}\ }\href
  {https://journals.aps.org/prl/abstract/10.1103/PhysRevLett.87.150601}
  {\bibfield  {journal} {\bibinfo  {journal} {Phys. Rev. Lett.}\ }\textbf
  {\bibinfo {volume} {87}},\ \bibinfo {pages} {150601} (\bibinfo {year}
  {2001})}\BibitemShut {NoStop}%
\bibitem [{\citenamefont {Bertini}\ \emph {et~al.}(2001)\citenamefont
  {Bertini}, \citenamefont {Sole}, \citenamefont {Gabrielli}, \citenamefont
  {Jona-Lasinio},\ and\ \citenamefont {Landim}}]{bertini01b}%
  \BibitemOpen
  \bibfield  {author} {\bibinfo {author} {\bibfnamefont {L.}~\bibnamefont
  {Bertini}}, \bibinfo {author} {\bibfnamefont {A.~De}\ \bibnamefont {Sole}},
  \bibinfo {author} {\bibfnamefont {D.}~\bibnamefont {Gabrielli}}, \bibinfo
  {author} {\bibfnamefont {G.}~\bibnamefont {Jona-Lasinio}}, \ and\ \bibinfo
  {author} {\bibfnamefont {C.}~\bibnamefont {Landim}},\ }\bibfield  {title}
  {\enquote {\bibinfo {title} {Large deviations for the boundary driven
  symmetric simple exclusion process},}\ }\href {\doibase
  10.1103/PhysRevLett.87.040601} {\bibfield  {journal} {\bibinfo  {journal}
  {Phys. Rev. Lett.}\ }\textbf {\bibinfo {volume} {87}},\ \bibinfo {pages}
  {040601} (\bibinfo {year} {2001})}\BibitemShut {NoStop}%
\bibitem [{\citenamefont {Bertini}\ \emph
  {et~al.}(2005{\natexlab{a}})\citenamefont {Bertini}, \citenamefont
  {Gabrielli},\ and\ \citenamefont {Lebowitz}}]{bertini05b}%
  \BibitemOpen
  \bibfield  {author} {\bibinfo {author} {\bibfnamefont {L.}~\bibnamefont
  {Bertini}}, \bibinfo {author} {\bibfnamefont {D.}~\bibnamefont {Gabrielli}},
  \ and\ \bibinfo {author} {\bibfnamefont {J.~L.}\ \bibnamefont {Lebowitz}},\
  }\bibfield  {title} {\enquote {\bibinfo {title} {Large deviations for a
  stochastic model of heat flow},}\ }\href
  {https://link.springer.com/article/10.1007/s10955-005-5527-2} {\bibfield
  {journal} {\bibinfo  {journal} {J. Stat. Phys.}\ }\textbf {\bibinfo {volume}
  {121}},\ \bibinfo {pages} {843} (\bibinfo {year}
  {2005}{\natexlab{a}})}\BibitemShut {NoStop}%
\bibitem [{\citenamefont {Bertini}\ \emph {et~al.}(2015)\citenamefont
  {Bertini}, \citenamefont {Sole}, \citenamefont {Gabrielli}, \citenamefont
  {Jona-Lasinio},\ and\ \citenamefont {Landim}}]{bertini15a}%
  \BibitemOpen
  \bibfield  {author} {\bibinfo {author} {\bibfnamefont {L.}~\bibnamefont
  {Bertini}}, \bibinfo {author} {\bibfnamefont {A.~De}\ \bibnamefont {Sole}},
  \bibinfo {author} {\bibfnamefont {D.}~\bibnamefont {Gabrielli}}, \bibinfo
  {author} {\bibfnamefont {G.}~\bibnamefont {Jona-Lasinio}}, \ and\ \bibinfo
  {author} {\bibfnamefont {C.}~\bibnamefont {Landim}},\ }\bibfield  {title}
  {\enquote {\bibinfo {title} {Macroscopic fluctuation theory},}\ }\href
  {http://journals.aps.org/rmp/abstract/10.1103/RevModPhys.87.593} {\bibfield
  {journal} {\bibinfo  {journal} {Rev. Mod. Phys.}\ }\textbf {\bibinfo {volume}
  {87}},\ \bibinfo {pages} {593--636} (\bibinfo {year} {2015})}\BibitemShut
  {NoStop}%
\bibitem [{\citenamefont {Touchette}(2009)}]{touchette09a}%
  \BibitemOpen
  \bibfield  {author} {\bibinfo {author} {\bibfnamefont {H.}~\bibnamefont
  {Touchette}},\ }\bibfield  {title} {\enquote {\bibinfo {title} {The large
  deviation approach to statistical mechanics},}\ }\href
  {http://dx.doi.org/10.1016/j.physrep.2009.05.002} {\bibfield  {journal}
  {\bibinfo  {journal} {Phys. Rep.}\ }\textbf {\bibinfo {volume} {478}},\
  \bibinfo {pages} {1} (\bibinfo {year} {2009})}\BibitemShut {NoStop}%
\bibitem [{\citenamefont {Bertini}\ \emph
  {et~al.}(2005{\natexlab{b}})\citenamefont {Bertini}, \citenamefont {Sole},
  \citenamefont {Gabrielli}, \citenamefont {Jona-Lasinio},\ and\ \citenamefont
  {Landim}}]{bertini05a}%
  \BibitemOpen
  \bibfield  {author} {\bibinfo {author} {\bibfnamefont {L.}~\bibnamefont
  {Bertini}}, \bibinfo {author} {\bibfnamefont {A.~De}\ \bibnamefont {Sole}},
  \bibinfo {author} {\bibfnamefont {D.}~\bibnamefont {Gabrielli}}, \bibinfo
  {author} {\bibfnamefont {G.}~\bibnamefont {Jona-Lasinio}}, \ and\ \bibinfo
  {author} {\bibfnamefont {C.}~\bibnamefont {Landim}},\ }\bibfield  {title}
  {\enquote {\bibinfo {title} {Current fluctuations in stochastic lattice
  gases},}\ }\href
  {http://journals.aps.org/prl/abstract/10.1103/PhysRevLett.94.030601}
  {\bibfield  {journal} {\bibinfo  {journal} {Phys. Rev. Lett.}\ }\textbf
  {\bibinfo {volume} {94}},\ \bibinfo {pages} {030601} (\bibinfo {year}
  {2005}{\natexlab{b}})}\BibitemShut {NoStop}%
\bibitem [{\citenamefont {Bertini}\ \emph {et~al.}(2006)\citenamefont
  {Bertini}, \citenamefont {Sole}, \citenamefont {Gabrielli}, \citenamefont
  {Jona-Lasinio},\ and\ \citenamefont {Landim}}]{bertini06a}%
  \BibitemOpen
  \bibfield  {author} {\bibinfo {author} {\bibfnamefont {L.}~\bibnamefont
  {Bertini}}, \bibinfo {author} {\bibfnamefont {A.~De}\ \bibnamefont {Sole}},
  \bibinfo {author} {\bibfnamefont {D.}~\bibnamefont {Gabrielli}}, \bibinfo
  {author} {\bibfnamefont {G.}~\bibnamefont {Jona-Lasinio}}, \ and\ \bibinfo
  {author} {\bibfnamefont {C.}~\bibnamefont {Landim}},\ }\bibfield  {title}
  {\enquote {\bibinfo {title} {Nonequilibrium current fluctuations in
  stochastic lattice gases},}\ }\href
  {http://link.springer.com/article/10.1007/s10955-006-9056-4} {\bibfield
  {journal} {\bibinfo  {journal} {J. Stat. Phys.}\ }\textbf {\bibinfo {volume}
  {123}},\ \bibinfo {pages} {237--276} (\bibinfo {year} {2006})}\BibitemShut
  {NoStop}%
\bibitem [{\citenamefont {Derrida}()}]{derrida07a}%
  \BibitemOpen
  \bibfield  {author} {\bibinfo {author} {\bibfnamefont {B.}~\bibnamefont
  {Derrida}},\ }\bibfield  {title} {\enquote {\bibinfo {title} {Non-equilibrium
  steady states: fluctuations and large deviations of the density and of the
  current},}\ }\href
  {http://iopscience.iop.org/article/10.1088/1742-5468/2007/07/P07023}
  {\bibinfo  {journal} {J. Stat. Mech. P07023 (2007)}\ }\BibitemShut {NoStop}%
\bibitem [{\citenamefont {Hurtado}\ \emph {et~al.}(2014)\citenamefont
  {Hurtado}, \citenamefont {Espigares}, \citenamefont {del Pozo},\ and\
  \citenamefont {Garrido}}]{hurtado14a}%
  \BibitemOpen
\bibfield  {journal} {  }\bibfield  {author} {\bibinfo {author} {\bibfnamefont
  {P.~I.}\ \bibnamefont {Hurtado}}, \bibinfo {author} {\bibfnamefont {C.~P.}\
  \bibnamefont {Espigares}}, \bibinfo {author} {\bibfnamefont {J.~J.}\
  \bibnamefont {del Pozo}}, \ and\ \bibinfo {author} {\bibfnamefont {P.~L.}\
  \bibnamefont {Garrido}},\ }\bibfield  {title} {\enquote {\bibinfo {title}
  {Thermodynamics of currents in nonequilibrium diffusive systems: theory and
  simulation},}\ }\href
  {http://link.springer.com/article/10.1007/s10955-013-0894-6} {\bibfield
  {journal} {\bibinfo  {journal} {J. Stat. Phys.}\ }\textbf {\bibinfo {volume}
  {154}},\ \bibinfo {pages} {214--264} (\bibinfo {year} {2014})}\BibitemShut
  {NoStop}%
\bibitem [{\citenamefont {P{\'e}rez-Espigares}\ and\ \citenamefont
  {Hurtado}(2019)}]{perez-espigares19a}%
  \BibitemOpen
  \bibfield  {author} {\bibinfo {author} {\bibfnamefont {C.}~\bibnamefont
  {P{\'e}rez-Espigares}}\ and\ \bibinfo {author} {\bibfnamefont {P.~I.}\
  \bibnamefont {Hurtado}},\ }\bibfield  {title} {\enquote {\bibinfo {title}
  {Sampling rare events across dynamical phase transitions},}\ }\href {\doibase
  10.1063/1.5091669} {\bibfield  {journal} {\bibinfo  {journal} {Chaos}\
  }\textbf {\bibinfo {volume} {29}},\ \bibinfo {pages} {083106} (\bibinfo
  {year} {2019})}\BibitemShut {NoStop}%
\bibitem [{\citenamefont {Hurtado}(2025)}]{hurtado25a}%
  \BibitemOpen
  \bibfield  {author} {\bibinfo {author} {\bibfnamefont {P.~I.}\ \bibnamefont
  {Hurtado}},\ }\href {https://arxiv.org/abs/2501.09629} {\enquote {\bibinfo
  {title} {Optimal paths and dynamical symmetry breaking in the current
  fluctuations of driven diffusive media},}\ } (\bibinfo {year} {2025}),\
  \Eprint {http://arxiv.org/abs/2501.09629} {arXiv:2501.09629
  [cond-mat.stat-mech]} \BibitemShut {NoStop}%
\bibitem [{\citenamefont {Derrida}\ and\ \citenamefont
  {Gerschenfeld}(2009)}]{derrida09a}%
  \BibitemOpen
  \bibfield  {author} {\bibinfo {author} {\bibfnamefont {B.}~\bibnamefont
  {Derrida}}\ and\ \bibinfo {author} {\bibfnamefont {A.}~\bibnamefont
  {Gerschenfeld}},\ }\bibfield  {title} {\enquote {\bibinfo {title} {Current
  {Fluctuations} in {One} {Dimensional} {Diffusive} {Systems} with a {Step}
  {Initial} {Density} profile},}\ }\href
  {http://link.springer.com/article/10.1007%2Fs10955-009-9830-1} {\bibfield
  {journal} {\bibinfo  {journal} {J. Stat. Phys.}\ }\textbf {\bibinfo {volume}
  {137}},\ \bibinfo {pages} {978--1000} (\bibinfo {year} {2009})}\BibitemShut
  {NoStop}%
\bibitem [{\citenamefont {Krapivsky}\ and\ \citenamefont
  {Meerson}(2012)}]{krapivsky12a}%
  \BibitemOpen
  \bibfield  {author} {\bibinfo {author} {\bibfnamefont {P.~L.}\ \bibnamefont
  {Krapivsky}}\ and\ \bibinfo {author} {\bibfnamefont {B.}~\bibnamefont
  {Meerson}},\ }\bibfield  {title} {\enquote {\bibinfo {title} {Fluctuations of
  current in nonstationary diffusive lattice gases},}\ }\href
  {http://journals.aps.org/pre/abstract/10.1103/PhysRevE.86.031106} {\bibfield
  {journal} {\bibinfo  {journal} {Phys. Rev. E}\ }\textbf {\bibinfo {volume}
  {86}},\ \bibinfo {pages} {031106} (\bibinfo {year} {2012})}\BibitemShut
  {NoStop}%
\bibitem [{\citenamefont {Meerson}\ and\ \citenamefont
  {Sasorov}()}]{meerson13a}%
  \BibitemOpen
  \bibfield  {author} {\bibinfo {author} {\bibfnamefont {B.}~\bibnamefont
  {Meerson}}\ and\ \bibinfo {author} {\bibfnamefont {P.~V.}\ \bibnamefont
  {Sasorov}},\ }\bibfield  {title} {\enquote {\bibinfo {title} {Extreme current
  fluctuations in a nonstationary stochastic heat flow},}\ }\href
  {http://iopscience.iop.org/article/10.1088/1742-5468/2013/12/P12011/meta}
  {\bibinfo  {journal} {J. Stat. Mech. P12011 (2013)}\ }\BibitemShut {NoStop}%
\bibitem [{\citenamefont {Bettelheim}\ \emph {et~al.}(2022)\citenamefont
  {Bettelheim}, \citenamefont {Smith},\ and\ \citenamefont
  {Meerson}}]{bettelheim22a}%
  \BibitemOpen
\bibfield  {journal} {  }\bibfield  {author} {\bibinfo {author} {\bibfnamefont
  {E.}~\bibnamefont {Bettelheim}}, \bibinfo {author} {\bibfnamefont {N.~R.}\
  \bibnamefont {Smith}}, \ and\ \bibinfo {author} {\bibfnamefont
  {B.}~\bibnamefont {Meerson}},\ }\bibfield  {title} {\enquote {\bibinfo
  {title} {Inverse scattering method solves the problem of full statistics of
  nonstationary heat transfer in the {Kipnis-Marchioro-Presutti} model},}\
  }\href {\doibase 10.1103/PhysRevLett.128.130602} {\bibfield  {journal}
  {\bibinfo  {journal} {Phys. Rev. Lett.}\ }\textbf {\bibinfo {volume} {128}},\
  \bibinfo {pages} {130602} (\bibinfo {year} {2022})}\BibitemShut {NoStop}%
\bibitem [{\citenamefont {P\'erez-Espigares}\ \emph {et~al.}(2016)\citenamefont
  {P\'erez-Espigares}, \citenamefont {Garrido},\ and\ \citenamefont
  {Hurtado}}]{perez-espigares16a}%
  \BibitemOpen
  \bibfield  {author} {\bibinfo {author} {\bibfnamefont {C.}~\bibnamefont
  {P\'erez-Espigares}}, \bibinfo {author} {\bibfnamefont {P.~L.}\ \bibnamefont
  {Garrido}}, \ and\ \bibinfo {author} {\bibfnamefont {P.~I.}\ \bibnamefont
  {Hurtado}},\ }\bibfield  {title} {\enquote {\bibinfo {title} {Weak additivity
  principle for current statistics in $d$-dimensions},}\ }\href
  {http://journals.aps.org/pre/abstract/10.1103/PhysRevE.93.040103} {\bibfield
  {journal} {\bibinfo  {journal} {Phys. Rev. E}\ }\textbf {\bibinfo {volume}
  {93}},\ \bibinfo {pages} {040103(R)} (\bibinfo {year} {2016})}\BibitemShut
  {NoStop}%
\bibitem [{\citenamefont {Tiz{\'o}n-Escamilla}\ \emph
  {et~al.}(2017)\citenamefont {Tiz{\'o}n-Escamilla}, \citenamefont {Hurtado},\
  and\ \citenamefont {Garrido}}]{tizon-escamilla17a}%
  \BibitemOpen
  \bibfield  {author} {\bibinfo {author} {\bibfnamefont {N.}~\bibnamefont
  {Tiz{\'o}n-Escamilla}}, \bibinfo {author} {\bibfnamefont {P.~I.}\
  \bibnamefont {Hurtado}}, \ and\ \bibinfo {author} {\bibfnamefont {P.~L.}\
  \bibnamefont {Garrido}},\ }\bibfield  {title} {\enquote {\bibinfo {title}
  {Structure of the optimal path to a fluctuation},}\ }\href
  {http://journals.aps.org/pre/pdf/10.1103/PhysRevE.95.002100} {\bibfield
  {journal} {\bibinfo  {journal} {Phys. Rev. E}\ }\textbf {\bibinfo {volume}
  {95}},\ \bibinfo {pages} {002100} (\bibinfo {year} {2017})}\BibitemShut
  {NoStop}%
\bibitem [{\citenamefont {Garrido}\ \emph {et~al.}(2019)\citenamefont
  {Garrido}, \citenamefont {Hurtado},\ and\ \citenamefont
  {Tiz{\'o}n-Escamilla}}]{garrido19a}%
  \BibitemOpen
  \bibfield  {author} {\bibinfo {author} {\bibfnamefont {P.~L.}\ \bibnamefont
  {Garrido}}, \bibinfo {author} {\bibfnamefont {P.~I.}\ \bibnamefont
  {Hurtado}}, \ and\ \bibinfo {author} {\bibfnamefont {N.}~\bibnamefont
  {Tiz{\'o}n-Escamilla}},\ }\bibfield  {title} {\enquote {\bibinfo {title}
  {Infinite family of universal profiles for heat current statistics in
  {F}ourier's law},}\ }\href
  {https://journals.aps.org/pre/abstract/10.1103/PhysRevE.99.022134} {\bibfield
   {journal} {\bibinfo  {journal} {Phys. Rev. E}\ }\textbf {\bibinfo {volume}
  {99}},\ \bibinfo {pages} {022134} (\bibinfo {year} {2019})}\BibitemShut
  {NoStop}%
\bibitem [{\citenamefont {Bodineau}\ and\ \citenamefont
  {Derrida}(2004)}]{bodineau04a}%
  \BibitemOpen
  \bibfield  {author} {\bibinfo {author} {\bibfnamefont {T.}~\bibnamefont
  {Bodineau}}\ and\ \bibinfo {author} {\bibfnamefont {B.}~\bibnamefont
  {Derrida}},\ }\bibfield  {title} {\enquote {\bibinfo {title} {Current
  fluctuations in nonequilibrium diffusive systems: {An} additivity
  principle},}\ }\href
  {http://journals.aps.org/prl/abstract/10.1103/PhysRevLett.92.180601}
  {\bibfield  {journal} {\bibinfo  {journal} {Phys. Rev. Lett.}\ }\textbf
  {\bibinfo {volume} {92}},\ \bibinfo {pages} {180601} (\bibinfo {year}
  {2004})}\BibitemShut {NoStop}%
\bibitem [{\citenamefont {Hurtado}\ and\ \citenamefont
  {Garrido}(2009)}]{hurtado09c}%
  \BibitemOpen
  \bibfield  {author} {\bibinfo {author} {\bibfnamefont {P.~I.}\ \bibnamefont
  {Hurtado}}\ and\ \bibinfo {author} {\bibfnamefont {P.~L.}\ \bibnamefont
  {Garrido}},\ }\bibfield  {title} {\enquote {\bibinfo {title} {Test of the
  {additivity} {principle} for {current} {fluctuations} in a {model} of {heat}
  conduction},}\ }\href
  {http://journals.aps.org/prl/abstract/10.1103/PhysRevLett.102.250601}
  {\bibfield  {journal} {\bibinfo  {journal} {Phys. Rev. Lett.}\ }\textbf
  {\bibinfo {volume} {102}},\ \bibinfo {pages} {250601} (\bibinfo {year}
  {2009})}\BibitemShut {NoStop}%
\bibitem [{\citenamefont {Hurtado}\ and\ \citenamefont
  {Garrido}(2010)}]{hurtado10a}%
  \BibitemOpen
  \bibfield  {author} {\bibinfo {author} {\bibfnamefont {P.~I.}\ \bibnamefont
  {Hurtado}}\ and\ \bibinfo {author} {\bibfnamefont {P.~L.}\ \bibnamefont
  {Garrido}},\ }\bibfield  {title} {\enquote {\bibinfo {title} {Large
  fluctuations of the macroscopic current in diffusive systems: {A} numerical
  test of the additivity principle},}\ }\href
  {http://journals.aps.org/pre/abstract/10.1103/PhysRevE.81.041102} {\bibfield
  {journal} {\bibinfo  {journal} {Phys. Rev. E}\ }\textbf {\bibinfo {volume}
  {81}},\ \bibinfo {pages} {041102} (\bibinfo {year} {2010})}\BibitemShut
  {NoStop}%
\bibitem [{\citenamefont {Bodineau}\ and\ \citenamefont
  {Derrida}(2006)}]{bodineau06a}%
  \BibitemOpen
  \bibfield  {author} {\bibinfo {author} {\bibfnamefont {T.}~\bibnamefont
  {Bodineau}}\ and\ \bibinfo {author} {\bibfnamefont {B.}~\bibnamefont
  {Derrida}},\ }\bibfield  {title} {\enquote {\bibinfo {title} {Current large
  deviations for asymmetric exclusion processes with open boundaries},}\ }\href
  {http://link.springer.com/article/10.1007/s10955-006-9048-4} {\bibfield
  {journal} {\bibinfo  {journal} {J. Stat. Phys.}\ }\textbf {\bibinfo {volume}
  {123}},\ \bibinfo {pages} {277--300} (\bibinfo {year} {2006})}\BibitemShut
  {NoStop}%
\bibitem [{\citenamefont {Hurtado}\ and\ \citenamefont
  {Garrido}(2011)}]{hurtado11a}%
  \BibitemOpen
  \bibfield  {author} {\bibinfo {author} {\bibfnamefont {P.~I.}\ \bibnamefont
  {Hurtado}}\ and\ \bibinfo {author} {\bibfnamefont {P.~L.}\ \bibnamefont
  {Garrido}},\ }\bibfield  {title} {\enquote {\bibinfo {title} {Spontaneous
  symmetry breaking at the fluctuating level},}\ }\href
  {http://journals.aps.org/prl/abstract/10.1103/PhysRevLett.107.180601}
  {\bibfield  {journal} {\bibinfo  {journal} {Phys. Rev. Lett.}\ }\textbf
  {\bibinfo {volume} {107}},\ \bibinfo {pages} {180601} (\bibinfo {year}
  {2011})}\BibitemShut {NoStop}%
\bibitem [{\citenamefont {P\'erez-Espigares}\ \emph {et~al.}(2013)\citenamefont
  {P\'erez-Espigares}, \citenamefont {Garrido},\ and\ \citenamefont
  {Hurtado}}]{perez-espigares13a}%
  \BibitemOpen
  \bibfield  {author} {\bibinfo {author} {\bibfnamefont {C.}~\bibnamefont
  {P\'erez-Espigares}}, \bibinfo {author} {\bibfnamefont {P.~L.}\ \bibnamefont
  {Garrido}}, \ and\ \bibinfo {author} {\bibfnamefont {P.~I.}\ \bibnamefont
  {Hurtado}},\ }\bibfield  {title} {\enquote {\bibinfo {title} {Dynamical phase
  transition for current statistics in a simple driven diffusive system},}\
  }\href {http://journals.aps.org/pre/abstract/10.1103/PhysRevE.87.032115}
  {\bibfield  {journal} {\bibinfo  {journal} {Phys. Rev. E}\ }\textbf {\bibinfo
  {volume} {87}},\ \bibinfo {pages} {032115} (\bibinfo {year}
  {2013})}\BibitemShut {NoStop}%
\bibitem [{\citenamefont {Zarfaty}\ and\ \citenamefont
  {Meerson}()}]{zarfaty16a}%
  \BibitemOpen
  \bibfield  {author} {\bibinfo {author} {\bibfnamefont {L.}~\bibnamefont
  {Zarfaty}}\ and\ \bibinfo {author} {\bibfnamefont {B.}~\bibnamefont
  {Meerson}},\ }\bibfield  {title} {\enquote {\bibinfo {title} {Statistics of
  large currents in the {Kipnis-Marchioro-Presutti} model in a ring
  geometry},}\ }\href
  {http://iopscience.iop.org/article/10.1088/1742-5468/2016/03/033304/meta}
  {\bibinfo  {journal} {J. Stat. Mech. P033304 (2016)}\ }\BibitemShut {NoStop}%
\bibitem [{\citenamefont {Evans}\ \emph {et~al.}(1993)\citenamefont {Evans},
  \citenamefont {Cohen},\ and\ \citenamefont {Morriss}}]{evans93a}%
  \BibitemOpen
\bibfield  {journal} {  }\bibfield  {author} {\bibinfo {author} {\bibfnamefont
  {D.~J.}\ \bibnamefont {Evans}}, \bibinfo {author} {\bibfnamefont {E.~G.~D.}\
  \bibnamefont {Cohen}}, \ and\ \bibinfo {author} {\bibfnamefont {G.~P.}\
  \bibnamefont {Morriss}},\ }\bibfield  {title} {\enquote {\bibinfo {title}
  {Probability of 2nd law violations in shearing steady-states},}\ }\href
  {http://journals.aps.org/prl/abstract/10.1103/PhysRevLett.71.2401} {\bibfield
   {journal} {\bibinfo  {journal} {Phys. Rev. Lett.}\ }\textbf {\bibinfo
  {volume} {71}},\ \bibinfo {pages} {2401--2404} (\bibinfo {year}
  {1993})}\BibitemShut {NoStop}%
\bibitem [{\citenamefont {Evans}\ and\ \citenamefont
  {Searles}(1994)}]{evans94a}%
  \BibitemOpen
  \bibfield  {author} {\bibinfo {author} {\bibfnamefont {D.J.}\ \bibnamefont
  {Evans}}\ and\ \bibinfo {author} {\bibfnamefont {D.J.}\ \bibnamefont
  {Searles}},\ }\bibfield  {title} {\enquote {\bibinfo {title} {Equilibrium
  microstates which generate second law violating steady-states},}\ }\href
  {https://journals.aps.org/pre/abstract/10.1103/PhysRevE.50.1645} {\bibfield
  {journal} {\bibinfo  {journal} {Phys. Rev. E}\ }\textbf {\bibinfo {volume}
  {50}},\ \bibinfo {pages} {1645} (\bibinfo {year} {1994})}\BibitemShut
  {NoStop}%
\bibitem [{\citenamefont {Gallavotti}\ and\ \citenamefont
  {Cohen}(1995{\natexlab{a}})}]{gallavotti95a}%
  \BibitemOpen
  \bibfield  {author} {\bibinfo {author} {\bibfnamefont {G.}~\bibnamefont
  {Gallavotti}}\ and\ \bibinfo {author} {\bibfnamefont {E.~G.~D.}\ \bibnamefont
  {Cohen}},\ }\bibfield  {title} {\enquote {\bibinfo {title} {Dynamical
  ensembles in nonequilibrium statistical mechanics},}\ }\href
  {http://journals.aps.org/prl/abstract/10.1103/PhysRevLett.74.2694} {\bibfield
   {journal} {\bibinfo  {journal} {Phys. Rev. Lett.}\ }\textbf {\bibinfo
  {volume} {74}},\ \bibinfo {pages} {2694} (\bibinfo {year}
  {1995}{\natexlab{a}})}\BibitemShut {NoStop}%
\bibitem [{\citenamefont {Gallavotti}\ and\ \citenamefont
  {Cohen}(1995{\natexlab{b}})}]{gallavotti95b}%
  \BibitemOpen
  \bibfield  {author} {\bibinfo {author} {\bibfnamefont {G.}~\bibnamefont
  {Gallavotti}}\ and\ \bibinfo {author} {\bibfnamefont {E.~G.~D.}\ \bibnamefont
  {Cohen}},\ }\bibfield  {title} {\enquote {\bibinfo {title} {Dynamical
  ensembles in stationary states},}\ }\href
  {http://link.springer.com/article/10.1007/BF02179860} {\bibfield  {journal}
  {\bibinfo  {journal} {J. Stat. Phys.}\ }\textbf {\bibinfo {volume} {80}},\
  \bibinfo {pages} {931} (\bibinfo {year} {1995}{\natexlab{b}})}\BibitemShut
  {NoStop}%
\bibitem [{\citenamefont {Kurchan}(1998)}]{kurchan98a}%
  \BibitemOpen
  \bibfield  {author} {\bibinfo {author} {\bibfnamefont {J.}~\bibnamefont
  {Kurchan}},\ }\bibfield  {title} {\enquote {\bibinfo {title} {Fluctuation
  theorem for stochastic dynamics},}\ }\href
  {http://iopscience.iop.org/article/10.1088/0305-4470/31/16/003/meta}
  {\bibfield  {journal} {\bibinfo  {journal} {J. Phys. A}\ }\textbf {\bibinfo
  {volume} {31}},\ \bibinfo {pages} {3719--3729} (\bibinfo {year}
  {1998})}\BibitemShut {NoStop}%
\bibitem [{\citenamefont {Lebowitz}\ and\ \citenamefont
  {Spohn}(1999)}]{lebowitz99a}%
  \BibitemOpen
  \bibfield  {author} {\bibinfo {author} {\bibfnamefont {J.~L.}\ \bibnamefont
  {Lebowitz}}\ and\ \bibinfo {author} {\bibfnamefont {H.}~\bibnamefont
  {Spohn}},\ }\bibfield  {title} {\enquote {\bibinfo {title} {A
  {Gallavotti-Cohen-type} symmetry in the large deviation functional for
  stochastic dynamics},}\ }\href
  {http://link.springer.com/article/10.1023%2FA%3A1004589714161} {\bibfield
  {journal} {\bibinfo  {journal} {J. Stat. Phys.}\ }\textbf {\bibinfo {volume}
  {95}},\ \bibinfo {pages} {333--365} (\bibinfo {year} {1999})}\BibitemShut
  {NoStop}%
\bibitem [{\citenamefont {Hurtado}\ \emph {et~al.}(2011)\citenamefont
  {Hurtado}, \citenamefont {P\'erez-Espigares}, \citenamefont {del Pozo},\ and\
  \citenamefont {Garrido}}]{hurtado11b}%
  \BibitemOpen
  \bibfield  {author} {\bibinfo {author} {\bibfnamefont {P.~I.}\ \bibnamefont
  {Hurtado}}, \bibinfo {author} {\bibfnamefont {C.}~\bibnamefont
  {P\'erez-Espigares}}, \bibinfo {author} {\bibfnamefont {J.~J.}\ \bibnamefont
  {del Pozo}}, \ and\ \bibinfo {author} {\bibfnamefont {P.~L.}\ \bibnamefont
  {Garrido}},\ }\bibfield  {title} {\enquote {\bibinfo {title} {Symmetries in
  fluctuations far from equilibrium},}\ }\href
  {http://www.pnas.org/content/108/19/7704.short} {\bibfield  {journal}
  {\bibinfo  {journal} {Proc. Natl. Acad. Sci. USA}\ }\textbf {\bibinfo
  {volume} {108}},\ \bibinfo {pages} {7704--7709} (\bibinfo {year}
  {2011})}\BibitemShut {NoStop}%
\bibitem [{\citenamefont {P\'erez-Espigares}\ \emph {et~al.}(2015)\citenamefont
  {P\'erez-Espigares}, \citenamefont {Redig},\ and\ \citenamefont
  {Giardin\`a}}]{perez-espigares15a}%
  \BibitemOpen
  \bibfield  {author} {\bibinfo {author} {\bibfnamefont {C.}~\bibnamefont
  {P\'erez-Espigares}}, \bibinfo {author} {\bibfnamefont {F.}~\bibnamefont
  {Redig}}, \ and\ \bibinfo {author} {\bibfnamefont {C.}~\bibnamefont
  {Giardin\`a}},\ }\bibfield  {title} {\enquote {\bibinfo {title} {Spatial
  fluctuation theorem},}\ }\href
  {http://iopscience.iop.org/article/10.1088/1751-8113/48/35/35FT01/meta}
  {\bibfield  {journal} {\bibinfo  {journal} {J. Phys. A}\ }\textbf {\bibinfo
  {volume} {48}},\ \bibinfo {pages} {35FT01} (\bibinfo {year}
  {2015})}\BibitemShut {NoStop}%
\bibitem [{\citenamefont {Giardin{\`a}}\ \emph {et~al.}(2007)\citenamefont
  {Giardin{\`a}}, \citenamefont {Kurchan},\ and\ \citenamefont
  {Redig}}]{giardina07a}%
  \BibitemOpen
  \bibfield  {author} {\bibinfo {author} {\bibfnamefont {C.}~\bibnamefont
  {Giardin{\`a}}}, \bibinfo {author} {\bibfnamefont {J.}~\bibnamefont
  {Kurchan}}, \ and\ \bibinfo {author} {\bibfnamefont {F.}~\bibnamefont
  {Redig}},\ }\bibfield  {title} {\enquote {\bibinfo {title} {Duality and exact
  correlations for a model of heat conduction},}\ }\href
  {https://aip.scitation.org/doi/abs/10.1063/1.2711373} {\bibfield  {journal}
  {\bibinfo  {journal} {J. Math. Phys.}\ }\textbf {\bibinfo {volume} {48}},\
  \bibinfo {pages} {033301} (\bibinfo {year} {2007})}\BibitemShut {NoStop}%
\bibitem [{\citenamefont {Giardin{\`a}}\ \emph {et~al.}(2009)\citenamefont
  {Giardin{\`a}}, \citenamefont {Kurchan}, \citenamefont {Redig},\ and\
  \citenamefont {Vafayi}}]{giardina09a}%
  \BibitemOpen
  \bibfield  {author} {\bibinfo {author} {\bibfnamefont {C.}~\bibnamefont
  {Giardin{\`a}}}, \bibinfo {author} {\bibfnamefont {J.}~\bibnamefont
  {Kurchan}}, \bibinfo {author} {\bibfnamefont {F.}~\bibnamefont {Redig}}, \
  and\ \bibinfo {author} {\bibfnamefont {K.}~\bibnamefont {Vafayi}},\
  }\bibfield  {title} {\enquote {\bibinfo {title} {Duality and hidden
  symmetries in interacting particle systems},}\ }\href
  {https://link.springer.com/article/10.1007/s10955-009-9716-2} {\bibfield
  {journal} {\bibinfo  {journal} {J. Stat. Phys.}\ }\textbf {\bibinfo {volume}
  {135}},\ \bibinfo {pages} {25} (\bibinfo {year} {2009})}\BibitemShut
  {NoStop}%
\bibitem [{\citenamefont {Carinci}\ \emph {et~al.}(2013)\citenamefont
  {Carinci}, \citenamefont {Giardina}, \citenamefont {Giberti},\ and\
  \citenamefont {Redig}}]{carinci13a}%
  \BibitemOpen
  \bibfield  {author} {\bibinfo {author} {\bibfnamefont {G.}~\bibnamefont
  {Carinci}}, \bibinfo {author} {\bibfnamefont {C.}~\bibnamefont {Giardina}},
  \bibinfo {author} {\bibfnamefont {C.}~\bibnamefont {Giberti}}, \ and\
  \bibinfo {author} {\bibfnamefont {F.}~\bibnamefont {Redig}},\ }\bibfield
  {title} {\enquote {\bibinfo {title} {Duality for stochastic models of
  transport},}\ }\href {\doibase 10.1007/s10955-013-0786-9} {\bibfield
  {journal} {\bibinfo  {journal} {J. Stat. Phys.}\ }\textbf {\bibinfo {volume}
  {152}},\ \bibinfo {pages} {657} (\bibinfo {year} {2013})}\BibitemShut
  {NoStop}%
\bibitem [{\citenamefont {Tailleur}\ \emph {et~al.}(2007)\citenamefont
  {Tailleur}, \citenamefont {Kurchan},\ and\ \citenamefont
  {Lecomte}}]{tailleur07a}%
  \BibitemOpen
  \bibfield  {author} {\bibinfo {author} {\bibfnamefont {J.}~\bibnamefont
  {Tailleur}}, \bibinfo {author} {\bibfnamefont {J.}~\bibnamefont {Kurchan}}, \
  and\ \bibinfo {author} {\bibfnamefont {V.}~\bibnamefont {Lecomte}},\
  }\bibfield  {title} {\enquote {\bibinfo {title} {Mapping nonequilibrium onto
  equilibrium: {The} macroscopic fluctuations of simple transport models},}\
  }\href {http://journals.aps.org/prl/abstract/10.1103/PhysRevLett.99.150602}
  {\bibfield  {journal} {\bibinfo  {journal} {Phys. Rev. Lett.}\ }\textbf
  {\bibinfo {volume} {99}},\ \bibinfo {pages} {150602} (\bibinfo {year}
  {2007})}\BibitemShut {NoStop}%
\bibitem [{\citenamefont {Hurtado}\ and\ \citenamefont
  {Krapivsky}(2012)}]{hurtado12a}%
  \BibitemOpen
  \bibfield  {author} {\bibinfo {author} {\bibfnamefont {P.~I.}\ \bibnamefont
  {Hurtado}}\ and\ \bibinfo {author} {\bibfnamefont {P.~L.}\ \bibnamefont
  {Krapivsky}},\ }\bibfield  {title} {\enquote {\bibinfo {title} {Compact waves
  in microscopic nonlinear diffusion},}\ }\href
  {http://journals.aps.org/pre/abstract/10.1103/PhysRevE.85.060103} {\bibfield
  {journal} {\bibinfo  {journal} {Phys. Rev. E}\ }\textbf {\bibinfo {volume}
  {85}},\ \bibinfo {pages} {060103} (\bibinfo {year} {2012})}\BibitemShut
  {NoStop}%
\bibitem [{\citenamefont {Guti{\'e}rrez-Ariza}\ and\ \citenamefont
  {Hurtado}(2019)}]{gutierrez-ariza19a}%
  \BibitemOpen
  \bibfield  {author} {\bibinfo {author} {\bibfnamefont {C.}~\bibnamefont
  {Guti{\'e}rrez-Ariza}}\ and\ \bibinfo {author} {\bibfnamefont {P.~I.}\
  \bibnamefont {Hurtado}},\ }\bibfield  {title} {\enquote {\bibinfo {title}
  {The kinetic exclusion process: a tale of two fields},}\ }\href
  {https://iopscience.iop.org/article/10.1088/1742-5468/ab4587} {\bibfield
  {journal} {\bibinfo  {journal} {J. Stat. Mech. 103203}\ } (\bibinfo {year}
  {2019})}\BibitemShut {NoStop}%
\bibitem [{\citenamefont {Prados}\ \emph {et~al.}(2011)\citenamefont {Prados},
  \citenamefont {Lasanta},\ and\ \citenamefont {Hurtado}}]{prados11a}%
  \BibitemOpen
  \bibfield  {author} {\bibinfo {author} {\bibfnamefont {A.}~\bibnamefont
  {Prados}}, \bibinfo {author} {\bibfnamefont {A.}~\bibnamefont {Lasanta}}, \
  and\ \bibinfo {author} {\bibfnamefont {P.~I.}\ \bibnamefont {Hurtado}},\
  }\bibfield  {title} {\enquote {\bibinfo {title} {Large fluctuations in driven
  dissipative media},}\ }\href
  {http://journals.aps.org/prl/abstract/10.1103/PhysRevLett.107.140601}
  {\bibfield  {journal} {\bibinfo  {journal} {Phys. Rev. Lett.}\ }\textbf
  {\bibinfo {volume} {107}},\ \bibinfo {pages} {140601} (\bibinfo {year}
  {2011})}\BibitemShut {NoStop}%
\bibitem [{\citenamefont {Prados}\ \emph {et~al.}(2012)\citenamefont {Prados},
  \citenamefont {Lasanta},\ and\ \citenamefont {Hurtado}}]{prados12a}%
  \BibitemOpen
  \bibfield  {author} {\bibinfo {author} {\bibfnamefont {A.}~\bibnamefont
  {Prados}}, \bibinfo {author} {\bibfnamefont {A.}~\bibnamefont {Lasanta}}, \
  and\ \bibinfo {author} {\bibfnamefont {P.~I.}\ \bibnamefont {Hurtado}},\
  }\bibfield  {title} {\enquote {\bibinfo {title} {Nonlinear driven diffusive
  systems with dissipation: {Fluctuating} hydrodynamics},}\ }\href
  {http://journals.aps.org/pre/abstract/10.1103/PhysRevE.86.031134} {\bibfield
  {journal} {\bibinfo  {journal} {Phys. Rev. E}\ }\textbf {\bibinfo {volume}
  {86}},\ \bibinfo {pages} {031134} (\bibinfo {year} {2012})}\BibitemShut
  {NoStop}%
\bibitem [{\citenamefont {Hurtado}\ \emph {et~al.}(2013)\citenamefont
  {Hurtado}, \citenamefont {Lasanta},\ and\ \citenamefont
  {Prados}}]{hurtado13a}%
  \BibitemOpen
  \bibfield  {author} {\bibinfo {author} {\bibfnamefont {P.~I.}\ \bibnamefont
  {Hurtado}}, \bibinfo {author} {\bibfnamefont {A.}~\bibnamefont {Lasanta}}, \
  and\ \bibinfo {author} {\bibfnamefont {A.}~\bibnamefont {Prados}},\
  }\bibfield  {title} {\enquote {\bibinfo {title} {Typical and rare
  fluctuations in nonlinear driven diffusive systems with dissipation},}\
  }\href {http://journals.aps.org/pre/abstract/10.1103/PhysRevE.88.022110}
  {\bibfield  {journal} {\bibinfo  {journal} {Phys. Rev. E}\ }\textbf {\bibinfo
  {volume} {88}},\ \bibinfo {pages} {022110} (\bibinfo {year}
  {2013})}\BibitemShut {NoStop}%
\bibitem [{\citenamefont {Lasanta}\ \emph {et~al.}(2015)\citenamefont
  {Lasanta}, \citenamefont {Manacorda}, \citenamefont {Prados},\ and\
  \citenamefont {Puglisi}}]{lasanta15a}%
  \BibitemOpen
  \bibfield  {author} {\bibinfo {author} {\bibfnamefont {A.}~\bibnamefont
  {Lasanta}}, \bibinfo {author} {\bibfnamefont {A.}~\bibnamefont {Manacorda}},
  \bibinfo {author} {\bibfnamefont {A.}~\bibnamefont {Prados}}, \ and\ \bibinfo
  {author} {\bibfnamefont {A.}~\bibnamefont {Puglisi}},\ }\bibfield  {title}
  {\enquote {\bibinfo {title} {Fluctuating hydrodynamics and mesoscopic effects
  of spatial correlations in dissipative systems with conserved momentum},}\
  }\href
  {https://iopscience.iop.org/article/10.1088/1367-2630/17/8/083039/meta}
  {\bibfield  {journal} {\bibinfo  {journal} {New J. Phys.}\ }\textbf {\bibinfo
  {volume} {17}},\ \bibinfo {pages} {083039} (\bibinfo {year}
  {2015})}\BibitemShut {NoStop}%
\bibitem [{\citenamefont {Manacorda}\ \emph {et~al.}(2016)\citenamefont
  {Manacorda}, \citenamefont {Plata}, \citenamefont {Lasanta}, \citenamefont
  {Puglisi},\ and\ \citenamefont {Prados}}]{manacorda16a}%
  \BibitemOpen
  \bibfield  {author} {\bibinfo {author} {\bibfnamefont {A.}~\bibnamefont
  {Manacorda}}, \bibinfo {author} {\bibfnamefont {C.~A.}\ \bibnamefont
  {Plata}}, \bibinfo {author} {\bibfnamefont {A.}~\bibnamefont {Lasanta}},
  \bibinfo {author} {\bibfnamefont {A.}~\bibnamefont {Puglisi}}, \ and\
  \bibinfo {author} {\bibfnamefont {A.}~\bibnamefont {Prados}},\ }\bibfield
  {title} {\enquote {\bibinfo {title} {Lattice models for granular-like
  velocity fields: hydrodynamic description},}\ }\href
  {https://link.springer.com/article/10.1007/s10955-016-1575-z} {\bibfield
  {journal} {\bibinfo  {journal} {J. Stat. Phys.}\ }\textbf {\bibinfo {volume}
  {164}},\ \bibinfo {pages} {810} (\bibinfo {year} {2016})}\BibitemShut
  {NoStop}%
\bibitem [{\citenamefont {Plata}\ \emph {et~al.}(2016)\citenamefont {Plata},
  \citenamefont {Manacorda}, \citenamefont {Lasanta}, \citenamefont {Puglisi},\
  and\ \citenamefont {Prados}}]{plata16a}%
  \BibitemOpen
  \bibfield  {author} {\bibinfo {author} {\bibfnamefont {C.~A.}\ \bibnamefont
  {Plata}}, \bibinfo {author} {\bibfnamefont {A.}~\bibnamefont {Manacorda}},
  \bibinfo {author} {\bibfnamefont {A.}~\bibnamefont {Lasanta}}, \bibinfo
  {author} {\bibfnamefont {A.}~\bibnamefont {Puglisi}}, \ and\ \bibinfo
  {author} {\bibfnamefont {A.}~\bibnamefont {Prados}},\ }\bibfield  {title}
  {\enquote {\bibinfo {title} {Lattice models for granular-like velocity
  fields: finite-size effects},}\ }\href
  {https://iopscience.iop.org/article/10.1088/1742-5468/2016/09/093203/meta}
  {\bibfield  {journal} {\bibinfo  {journal} {J. Stat. Mech. 093203}\ }
  (\bibinfo {year} {2016})}\BibitemShut {NoStop}%
\bibitem [{\citenamefont {Wilczek}(2012)}]{wilczek12a}%
  \BibitemOpen
  \bibfield  {author} {\bibinfo {author} {\bibfnamefont {F.}~\bibnamefont
  {Wilczek}},\ }\bibfield  {title} {\enquote {\bibinfo {title} {Quantum time
  crystals},}\ }\href {\doibase 10.1103/PhysRevLett.109.160401} {\bibfield
  {journal} {\bibinfo  {journal} {Phys. Rev. Lett.}\ }\textbf {\bibinfo
  {volume} {109}},\ \bibinfo {pages} {160401} (\bibinfo {year}
  {2012})}\BibitemShut {NoStop}%
\bibitem [{\citenamefont {Zakrzewski}(2012)}]{zakrzewski12a}%
  \BibitemOpen
  \bibfield  {author} {\bibinfo {author} {\bibfnamefont {J.}~\bibnamefont
  {Zakrzewski}},\ }\bibfield  {title} {\enquote {\bibinfo {title} {Crystals of
  time},}\ }\href
  {https://physics.aps.org/articles/v5/116?__hstc=13887208.23fad615fbe7758f631f836e863ee2ed.1473638400048.1473638400050.1473638400051.2&__hssc=13887208.1.1473638400051&__hsfp=1773666937}
  {\bibfield  {journal} {\bibinfo  {journal} {Physics}\ }\textbf {\bibinfo
  {volume} {5}},\ \bibinfo {pages} {116} (\bibinfo {year} {2012})}\BibitemShut
  {NoStop}%
\bibitem [{\citenamefont {Sacha}\ and\ \citenamefont
  {Zakrzewski}(2018)}]{sacha18a}%
  \BibitemOpen
  \bibfield  {author} {\bibinfo {author} {\bibfnamefont {K.}~\bibnamefont
  {Sacha}}\ and\ \bibinfo {author} {\bibfnamefont {J.}~\bibnamefont
  {Zakrzewski}},\ }\bibfield  {title} {\enquote {\bibinfo {title} {Time
  crystals: a review},}\ }\href {\doibase 10.1088/1361-6633/aa8b38} {\bibfield
  {journal} {\bibinfo  {journal} {Rep. Prog. Phys.}\ }\textbf {\bibinfo
  {volume} {81}},\ \bibinfo {pages} {016401} (\bibinfo {year}
  {2018})}\BibitemShut {NoStop}%
\bibitem [{\citenamefont {Sacha}(2020)}]{sacha20a}%
  \BibitemOpen
  \bibfield  {author} {\bibinfo {author} {\bibfnamefont {K.}~\bibnamefont
  {Sacha}},\ }\href@noop {} {\emph {\bibinfo {title} {Time crystals}}},\
  \bibinfo {series} {Springer Series on Atomic, Optical, and Plasma Physics},
  Vol.\ \bibinfo {volume} {114}\ (\bibinfo  {publisher} {Springer},\ \bibinfo
  {year} {2020})\BibitemShut {NoStop}%
\bibitem [{\citenamefont {Hurtado-Guti\'errez}\ \emph
  {et~al.}(2020)\citenamefont {Hurtado-Guti\'errez}, \citenamefont {Carollo},
  \citenamefont {P\'erez-Espigares},\ and\ \citenamefont
  {Hurtado}}]{hurtado-gutierrez20a}%
  \BibitemOpen
  \bibfield  {author} {\bibinfo {author} {\bibfnamefont {R.}~\bibnamefont
  {Hurtado-Guti\'errez}}, \bibinfo {author} {\bibfnamefont {F.}~\bibnamefont
  {Carollo}}, \bibinfo {author} {\bibfnamefont {C.}~\bibnamefont
  {P\'erez-Espigares}}, \ and\ \bibinfo {author} {\bibfnamefont {P.~I.}\
  \bibnamefont {Hurtado}},\ }\bibfield  {title} {\enquote {\bibinfo {title}
  {Building continuous time crystals from rare events},}\ }\href {\doibase
  10.1103/PhysRevLett.125.160601} {\bibfield  {journal} {\bibinfo  {journal}
  {Phys. Rev. Lett.}\ }\textbf {\bibinfo {volume} {125}},\ \bibinfo {pages}
  {160601} (\bibinfo {year} {2020})}\BibitemShut {NoStop}%
\bibitem [{\citenamefont {Hurtado-Guti\'errez}\ \emph
  {et~al.}(2023)\citenamefont {Hurtado-Guti\'errez}, \citenamefont {Hurtado},\
  and\ \citenamefont {P\'erez-Espigares}}]{hurtado-gutierrez23a}%
  \BibitemOpen
  \bibfield  {author} {\bibinfo {author} {\bibfnamefont {R.}~\bibnamefont
  {Hurtado-Guti\'errez}}, \bibinfo {author} {\bibfnamefont {P.~I.}\
  \bibnamefont {Hurtado}}, \ and\ \bibinfo {author} {\bibfnamefont
  {C.}~\bibnamefont {P\'erez-Espigares}},\ }\bibfield  {title} {\enquote
  {\bibinfo {title} {Spectral signatures of symmetry-breaking dynamical phase
  transitions},}\ }\href
  {https://journals.aps.org/pre/abstract/10.1103/PhysRevE.108.014107}
  {\bibfield  {journal} {\bibinfo  {journal} {Phys. Rev. E}\ }\textbf {\bibinfo
  {volume} {108}},\ \bibinfo {pages} {014107} (\bibinfo {year}
  {2023})}\BibitemShut {NoStop}%
\bibitem [{\citenamefont {Hurtado-Guti\'errez}\ \emph
  {et~al.}(2025{\natexlab{a}})\citenamefont {Hurtado-Guti\'errez},
  \citenamefont {P\'erez-Espigares},\ and\ \citenamefont
  {Hurtado}}]{hurtado-gutierrez25a}%
  \BibitemOpen
  \bibfield  {author} {\bibinfo {author} {\bibfnamefont {R.}~\bibnamefont
  {Hurtado-Guti\'errez}}, \bibinfo {author} {\bibfnamefont {C.}~\bibnamefont
  {P\'erez-Espigares}}, \ and\ \bibinfo {author} {\bibfnamefont {P.~I.}\
  \bibnamefont {Hurtado}},\ }\bibfield  {title} {\enquote {\bibinfo {title}
  {Programmable time crystals from higher-order packing fields},}\ }\href
  {https://journals.aps.org/pre/accepted/5b07bY02H622ed0641fc23a87bd1d4668d7d0af6f}
  {\bibfield  {journal} {\bibinfo  {journal} {Phys. Rev. E}\ }\textbf {\bibinfo
  {volume} {111}},\ \bibinfo {pages} {034119} (\bibinfo {year}
  {2025}{\natexlab{a}})}\BibitemShut {NoStop}%
\bibitem [{\citenamefont {Hurtado-Guti\'errez}\ \emph
  {et~al.}(2025{\natexlab{b}})\citenamefont {Hurtado-Guti\'errez},
  \citenamefont {P\'erez-Espigares},\ and\ \citenamefont
  {Hurtado}}]{hurtado-gutierrez25b}%
  \BibitemOpen
  \bibfield  {author} {\bibinfo {author} {\bibfnamefont {R.}~\bibnamefont
  {Hurtado-Guti\'errez}}, \bibinfo {author} {\bibfnamefont {C.}~\bibnamefont
  {P\'erez-Espigares}}, \ and\ \bibinfo {author} {\bibfnamefont {P.~I.}\
  \bibnamefont {Hurtado}},\ }\bibfield  {title} {\enquote {\bibinfo {title}
  {Critical behavior of a programmable time-crystal lattice gas},}\ }\href
  {\doibase 10.1103/rx8l-f4ql} {\bibfield  {journal} {\bibinfo  {journal}
  {Phys. Rev. E}\ }\textbf {\bibinfo {volume} {112}},\ \bibinfo {pages}
  {044135} (\bibinfo {year} {2025}{\natexlab{b}})}\BibitemShut {NoStop}%
\bibitem [{\citenamefont {Carinci}\ \emph {et~al.}(2016)\citenamefont
  {Carinci}, \citenamefont {Giardina}, \citenamefont {Redig},\ and\
  \citenamefont {Sasamoto}}]{carinci16a}%
  \BibitemOpen
  \bibfield  {author} {\bibinfo {author} {\bibfnamefont {G.}~\bibnamefont
  {Carinci}}, \bibinfo {author} {\bibfnamefont {C.}~\bibnamefont {Giardina}},
  \bibinfo {author} {\bibfnamefont {F.}~\bibnamefont {Redig}}, \ and\ \bibinfo
  {author} {\bibfnamefont {T.}~\bibnamefont {Sasamoto}},\ }\bibfield  {title}
  {\enquote {\bibinfo {title} {Asymmetric stochastic transport models with
  $\mathcal{U}_q(\mathfrak{su}(1,1))$ symmetry},}\ }\href {\doibase
  https://doi.org/10.1007/s10955-016-1473-4} {\bibfield  {journal} {\bibinfo
  {journal} {J. Stat. Phys.}\ }\textbf {\bibinfo {volume} {163}},\ \bibinfo
  {pages} {239} (\bibinfo {year} {2016})}\BibitemShut {NoStop}%
\bibitem [{\citenamefont {Spohn}(2012)}]{spohn12a}%
  \BibitemOpen
  \bibfield  {author} {\bibinfo {author} {\bibfnamefont {H.}~\bibnamefont
  {Spohn}},\ }\href {https://www.springer.com/la/book/9783642843730} {\emph
  {\bibinfo {title} {{Large Scale Dynamics of Interacting Particles}}}},\
  {Theoretical and Mathematical Physics}\ (\bibinfo  {publisher} {Springer
  Berlin Heidelberg},\ \bibinfo {year} {2012})\BibitemShut {NoStop}%
\bibitem [{\citenamefont {Van~Kampen}(2011)}]{kampen11a}%
  \BibitemOpen
  \bibfield  {author} {\bibinfo {author} {\bibfnamefont {N.G.}\ \bibnamefont
  {Van~Kampen}},\ }\href {https://books.google.es/books?id=N6II-6HlPxEC} {\emph
  {\bibinfo {title} {Stochastic {Processes} in {Physics} and {Chemistry}}}},\
  North-{Holland} {Personal} {Library}\ (\bibinfo  {publisher} {Elsevier
  Science},\ \bibinfo {year} {2011})\BibitemShut {NoStop}%
\bibitem [{\citenamefont {{De Groot}}\ and\ \citenamefont
  {Mazur}(2013)}]{de-groot13a}%
  \BibitemOpen
  \bibfield  {author} {\bibinfo {author} {\bibfnamefont {S.R.}\ \bibnamefont
  {{De Groot}}}\ and\ \bibinfo {author} {\bibfnamefont {P.}~\bibnamefont
  {Mazur}},\ }\href@noop {} {\emph {\bibinfo {title} {{Non-Equilibrium
  Thermodynamics}}}},\ {Dover Books on Physics}\ (\bibinfo  {publisher} {Dover
  Publications, New York},\ \bibinfo {year} {2013})\BibitemShut {NoStop}%
\bibitem [{\citenamefont {Ortiz~de Zarate}\ and\ \citenamefont
  {Sengers}(2006)}]{zarate06a}%
  \BibitemOpen
  \bibfield  {author} {\bibinfo {author} {\bibfnamefont {J.M.}\ \bibnamefont
  {Ortiz~de Zarate}}\ and\ \bibinfo {author} {\bibfnamefont {J.V.}\
  \bibnamefont {Sengers}},\ }\href@noop {} {\emph {\bibinfo {title}
  {Hydrodynamic fluctuations in fluids and fluid mixtures}}}\ (\bibinfo
  {publisher} {Elsevier, Amsterdam},\ \bibinfo {year} {2006})\BibitemShut
  {NoStop}%
\bibitem [{Note1()}]{Note1}%
  \BibitemOpen
  \bibinfo {note} {In particular, while the convergence of the
  integrals~\protect \eqref {diff}-\protect \eqref {mobi} for the diffusivity
  and mobility transport coefficients in terms of the microscopic collision
  kernel $\pi (\nu )$ demands $\beta >-4$, the convergence of the
  energy-dissipation coefficient for the granular or dissipative generalization
  of the nonlinear KMP model impose the stiffer restriction $\beta >-3$, see
  Ref. \cite {prados12a}.}\BibitemShut {Stop}%
\bibitem [{\citenamefont {Bodineau}\ and\ \citenamefont
  {Derrida}(2005)}]{bodineau05a}%
  \BibitemOpen
  \bibfield  {author} {\bibinfo {author} {\bibfnamefont {T.}~\bibnamefont
  {Bodineau}}\ and\ \bibinfo {author} {\bibfnamefont {B.}~\bibnamefont
  {Derrida}},\ }\bibfield  {title} {\enquote {\bibinfo {title} {Distribution of
  current in nonequilibrium diffusive systems and phase transitions},}\ }\href
  {http://journals.aps.org/pre/abstract/10.1103/PhysRevE.72.066110} {\bibfield
  {journal} {\bibinfo  {journal} {Phys. Rev. E}\ }\textbf {\bibinfo {volume}
  {72}},\ \bibinfo {pages} {066110} (\bibinfo {year} {2005})}\BibitemShut
  {NoStop}%
\bibitem [{\citenamefont {Doob}(1957)}]{doob57a}%
  \BibitemOpen
  \bibfield  {author} {\bibinfo {author} {\bibfnamefont {J.~L.}\ \bibnamefont
  {Doob}},\ }\bibfield  {title} {\enquote {\bibinfo {title} {Conditional
  {B}rownian motion and the boundary limits of harmonic functions},}\ }\href
  {https://eudml.org/doc/86928} {\bibfield  {journal} {\bibinfo  {journal}
  {Bull. Soc. Math. Fr.}\ }\textbf {\bibinfo {volume} {85}},\ \bibinfo {pages}
  {431} (\bibinfo {year} {1957})}\BibitemShut {NoStop}%
\bibitem [{\citenamefont {Chetrite}\ and\ \citenamefont
  {Touchette}(2015{\natexlab{a}})}]{chetrite15a}%
  \BibitemOpen
  \bibfield  {author} {\bibinfo {author} {\bibfnamefont {R.}~\bibnamefont
  {Chetrite}}\ and\ \bibinfo {author} {\bibfnamefont {H.}~\bibnamefont
  {Touchette}},\ }\bibfield  {title} {\enquote {\bibinfo {title} {Variational
  and optimal control representations of conditioned and driven processes},}\
  }\href
  {http://iopscience.iop.org/article/10.1088/1742-5468/2015/12/P12001/meta}
  {\bibfield  {journal} {\bibinfo  {journal} {J. Stat. Mech. P12001}\ }
  (\bibinfo {year} {2015}{\natexlab{a}})}\BibitemShut {NoStop}%
\bibitem [{\citenamefont {Chetrite}\ and\ \citenamefont
  {Touchette}(2015{\natexlab{b}})}]{chetrite15b}%
  \BibitemOpen
  \bibfield  {author} {\bibinfo {author} {\bibfnamefont {R.}~\bibnamefont
  {Chetrite}}\ and\ \bibinfo {author} {\bibfnamefont {H.}~\bibnamefont
  {Touchette}},\ }\bibfield  {title} {\enquote {\bibinfo {title}
  {Nonequilibrium {Markov} processes conditioned on large deviations},}\ }\href
  {https://link.springer.com/article/10.1007%2Fs00023-014-0375-8} {\bibfield
  {journal} {\bibinfo  {journal} {Ann. Henri Poincare}\ }\textbf {\bibinfo
  {volume} {16}},\ \bibinfo {pages} {2005} (\bibinfo {year}
  {2015}{\natexlab{b}})}\BibitemShut {NoStop}%
\bibitem [{\citenamefont {Carollo}\ \emph {et~al.}(2018)\citenamefont
  {Carollo}, \citenamefont {Garrahan}, \citenamefont {Lesanovsky},\ and\
  \citenamefont {P\'erez-Espigares}}]{carollo18b}%
  \BibitemOpen
  \bibfield  {author} {\bibinfo {author} {\bibfnamefont {F.}~\bibnamefont
  {Carollo}}, \bibinfo {author} {\bibfnamefont {J.~P.}\ \bibnamefont
  {Garrahan}}, \bibinfo {author} {\bibfnamefont {I.}~\bibnamefont
  {Lesanovsky}}, \ and\ \bibinfo {author} {\bibfnamefont {C.}~\bibnamefont
  {P\'erez-Espigares}},\ }\bibfield  {title} {\enquote {\bibinfo {title}
  {Making rare events typical in {M}arkovian open quantum systems},}\ }\href
  {\doibase 10.1103/PhysRevA.98.010103} {\bibfield  {journal} {\bibinfo
  {journal} {Phys. Rev. A}\ }\textbf {\bibinfo {volume} {98}},\ \bibinfo
  {pages} {010103} (\bibinfo {year} {2018})}\BibitemShut {NoStop}%
\bibitem [{\citenamefont {Simon}(2009)}]{simon09a}%
  \BibitemOpen
  \bibfield  {author} {\bibinfo {author} {\bibfnamefont {D.}~\bibnamefont
  {Simon}},\ }\bibfield  {title} {\enquote {\bibinfo {title} {Construction of a
  coordinate {B}ethe ansatz for the asymmetric simple exclusion process with
  open boundaries},}\ }\href
  {https://doi.org/10.1088%2F1742-5468%2F2009%2F07%2Fp07017} {\bibfield
  {journal} {\bibinfo  {journal} {J. Stat. Mech. P07017}\ } (\bibinfo {year}
  {2009})}\BibitemShut {NoStop}%
\bibitem [{\citenamefont {Jack}\ and\ \citenamefont {Sollich}(2010)}]{jack10a}%
  \BibitemOpen
  \bibfield  {author} {\bibinfo {author} {\bibfnamefont {R.~L.}\ \bibnamefont
  {Jack}}\ and\ \bibinfo {author} {\bibfnamefont {P.}~\bibnamefont {Sollich}},\
  }\bibfield  {title} {\enquote {\bibinfo {title} {Large deviations and
  ensembles of trajectories in stochastic models},}\ }\href {\doibase
  10.1143/PTPS.184.304} {\bibfield  {journal} {\bibinfo  {journal} {Prog.
  Theor. Phys. Supp.}\ }\textbf {\bibinfo {volume} {184}},\ \bibinfo {pages}
  {304} (\bibinfo {year} {2010})}\BibitemShut {NoStop}%
\bibitem [{\citenamefont {Popkov}\ \emph {et~al.}(2010)\citenamefont {Popkov},
  \citenamefont {Sch{\"u}tz},\ and\ \citenamefont {Simon}}]{popkov10a}%
  \BibitemOpen
  \bibfield  {author} {\bibinfo {author} {\bibfnamefont {V.}~\bibnamefont
  {Popkov}}, \bibinfo {author} {\bibfnamefont {G.~M.}\ \bibnamefont
  {Sch{\"u}tz}}, \ and\ \bibinfo {author} {\bibfnamefont {D.}~\bibnamefont
  {Simon}},\ }\bibfield  {title} {\enquote {\bibinfo {title} {{ASEP} on a ring
  conditioned on enhanced flux},}\ }\href
  {http://dx.doi.org/10.1088/1742-5468/2010/10/P10007} {\bibfield  {journal}
  {\bibinfo  {journal} {J. Stat. Mech. P10007}\ } (\bibinfo {year}
  {2010})}\BibitemShut {NoStop}%
\end{thebibliography}%

\end{document}